\documentclass[prd,aps,amssymb,showpacs,twocolumn]{revtex4}
\usepackage{amsfonts}
\usepackage{amsmath}
\usepackage{mathrsfs}
\usepackage{graphicx}
\usepackage{color}

\renewcommand\a{\alpha}
\renewcommand\b{\beta}
\renewcommand\d{\delta}
\renewcommand\k{\kappa}
\renewcommand\l{\lambda}
\renewcommand\r{\rho}

\renewcommand\t{\tau}

\renewcommand\c{\chi}

\renewcommand\o{\omega}
\newcommand\e{\epsilon}
\newcommand\g{\gamma}
\newcommand\z{\zeta}
\newcommand\m{\mu}
\newcommand\n{\nu}

\newcommand\p{\pi}
\newcommand\h{\theta}
\newcommand\s{\sigma}
\newcommand\f{\phi}
\newcommand\w{\eta}
\newcommand\ve{\varepsilon}

\newcommand\vf{\varphi}

\renewcommand\O{\Omega}

\newcommand\D{\Delta}
\newcommand\G{\Gamma}
\newcommand\F{\Phi}

\newcommand{\fig}[1]{Fig.~\ref{#1}}
\newcommand{\eq}[1]{Eq.~(\ref{#1})}
\newcommand{\eqs}[2]{Eqs.~(\ref{#1})-(\ref{#2})}

\newcommand\lb{\left(}
\newcommand\rb{\right)}
\newcommand\ls{\left[}
\newcommand\rs{\right]}

\newcommand\ra{\rightarrow}

\newcommand{\non}{\nonumber\\}
\newcommand\pt{\partial}
\newcommand{\diag}{{\rm{diag}}}
\newcommand{\fm}{{\rm fm}}

\newcommand{\re}{{\rm{Re}}}

\newcommand{\cm}{{\cal M}}

\newcommand{\bv}{{\mathbf v}}

\newcommand{\bz}{{\mathbf z}}
\newcommand{\br}{{\mathbf r}}

\newcommand{\bk}{{\mathbf k}}

\newcommand{\ot}{{\tilde \o}}
\newcommand{\pperp}{{P_\perp}}
\newcommand{\ppara}{{P_\parallel}}
\newcommand{\zperp}{{\z_\perp}}
\newcommand{\zpara}{{\z_\parallel}}
\newcommand{\cperp}{{C_\perp}}
\newcommand{\cpara}{{C_\parallel}}

\newcommand{\ie}{\emph{i.e.}}

\newcommand{\eg}{\emph{e.g.}}

\begin{document}

\title{Anisotropic Hydrodynamics, Bulk Viscosities and R-Modes of Strange Quark Stars with Strong Magnetic Fields}
\author{Xu-Guang Huang$^{1,2,3,4}$, Mei Huang$^3$, Dirk H. Rischke$^{1,2}$, and Armen Sedrakian$^2$}
\affiliation{$^1$ Frankfurt Institute for Advanced Studies, D-60438 Frankfurt am Main, Germany\\
$^2$ Institut f\"ur Theoretische Physik, J. W.
Goethe-Universit\"at,D-60438
Frankfurt am Main, Germany\\
$^3$Institute of High Energy Physics, Chinese Academy of Sciences,
Beijing 100039, China\\
$^4$Physics Department, Tsinghua University, Beijing 100084, China}

\date{\today}

\begin{abstract}
In strong magnetic fields the transport coefficients of strange
quark matter become anisotropic. We determine the general form of
the complete set of transport coefficients in the presence of a
strong magnetic field. By using a local linear response method, we
calculate explicitly the bulk viscosities $\zperp$ and $\zpara$
transverse and parallel to the $B$-field respectively, which arise
due to the non-leptonic weak processes $u+s\leftrightarrow u+d$. We
find that for magnetic fields $B<10^{17}$ G, the dependence of
$\zperp$ and $\zpara$ on the field is weak, and they can be
approximated by the bulk viscosity for zero magnetic field. For
fields $B>10^{18}$ G, the dependence of both $\zperp$ and $\zpara$
on the field is strong, and they exhibit de Haas-van Alphen-type
oscillations. With increasing magnetic field, the amplitude of these
oscillations increases, which eventually leads to negative $\zperp$
in some regions of parameter space. We show that the change of sign
of $\zperp$ signals a hydrodynamic instability. As an application,
we discuss the effects of the new bulk viscosities on the r-mode
instability in rotating strange quark stars.  We find that the
instability region in strange quark stars is affected when the
magnetic fields exceeds the value $B= 10^{17}$ G. For fields which
are larger by an order of magnitude, the instability region is
significantly enlarged, making magnetized strange stars more
susceptible to $r$-mode instability than their unmagnetized
counterparts.
\end{abstract}

\maketitle

%%%%%%%%%%%%%%%%%%%%%%%%%%%%%%%%%%%%%%%%%%%%%%%%%%%%%%%%%%%%%%%%%%%%%%%
\section {Introduction}\label{1}
%%%%%%%%%%%%%%%%%%%%%%%%%%%%%%%%%%%%%%%%%%%%%%%%%%%%%%%%%%%%%%%%%%%%%%%
Neutron stars provide a natural laboratory to study extremely dense
matter. In the interiors of such stars, the density can reach up to
several times the nuclear saturation density, $n_0\simeq 0.16
\,\fm^{-3}$. At such high densities quarks could be squeezed out of
nucleons to form quark
matter~\cite{Ivanenko:1965dg,Itoh:1970uw,Collins:1974ky}. The true
ground state of dense quark matter at high densities and low
temperatures remains an open problem due to the difficulty of
solving non-perturbative  quantum chromodynamics (QCD). It has been
suggested that strange quark matter that consists of comparable
numbers of $u$, $d$, and $s$ quarks may be the stable ground state
of normal quark matter~\cite{Witten:1984rs}. This led to the
conjecture that the family of compact stars may have members
consisting entirely of quark matter (so-called strange stars) and/or
members featuring quark cores surrounded by a hadronic shell (hybrid
stars)~\cite{Weber:2004kj}.

Observationally, it is very challenging to distinguish the various
types of compact objects, such as the strange stars, hybrid stars,
and ordinary neutron stars. Their early cooling behavior is
dominated by neutrino emission which is a useful probe of the
internal composition of compact stars. Thus, cooling simulations
provide an effective test of the nature of compact
stars~\cite{Iwamoto:1980eb,Blaschke:1999qx,Blaschke:2000dy,Schafer:2004jp,Alford:2004zr,Jaikumar:2005hy,Schmitt:2005wg,Anglani:2006br,Huang:2007jw,Huang:2007cr,Kouvaris:2009pc}.
However, many theoretical uncertainties and the current amount of
data on the surface temperatures of neutron stars leave sufficient
room for
speculations~\cite{Yakovlev:2004iq,Sedrakian:2006mq,Blaschke:2006gd}.
Another useful avenue for testing the internal structure and
composition of compact stars is astro-seismology, i.e., the study of
the phenomena related to stellar
vibrations~\cite{gr-qc/9706075,gr-qc/9706073,gr-qc/0010102,astro-ph/0101136,arXiv:0806.1005,Knippel:2009st}.
In particular, there are a number of instabilities which are
associated with the oscillations of rotating stars. Here we will be
concerned with the so-called r-mode instability [see
Refs.~\cite{gr-qc/0010102,astro-ph/0101136} for reviews]. This
instability is known to limit the angular velocity of rapidly
rotating compact stars. The r-mode and related instabilities in
rotating neutron stars are damped by the shear and bulk viscosities
of matter, therefore these are important ingredients of
theoretically modelling rapidly rotating stars. Such models and
their microscopic input can then be constrained via the observations
of rapidly rotating pulsars, such as the Crab pulsar and the
millisecond pulsars.

For quark matter in chemical equilibrium, the shear viscosity is
dominated by strong interactions between quarks. The bulk viscosity,
however, is dominated by flavor-changing weak processes, whereas
strong interactions play a secondary role. For normal
(non-superconducting) strange quark matter, the bulk viscosity is
dominated by the non-leptonic
process~\cite{Wang:1985tg,Sawyer:1989uy,Madsen:1992sx,Zheng:2002jq,Xiaoping:2005js}
\begin{subequations}
\begin{eqnarray}
\label{weak1} u+s\ra u+d,
\end{eqnarray}
\begin{eqnarray}
\label{weak2} u+d\ra u+s,
\end{eqnarray}
\end{subequations}
since the contributions of the leptonic processes
$u+e\leftrightarrow d+\n$ and $u+e\leftrightarrow s+\n$ are
suppressed due to much smaller phase spaces. The bulk viscosity of
various phases of quark matter has been studied extensively, see
Refs.~\cite{Wang:1985tg,Sawyer:1989uy,Madsen:1992sx,Zheng:2002jq,Xiaoping:2005js,Anand:1999bj,Sad:2006qv,Alford:2006gy,Sad:2007ud,Dong:2007mb,Alford:2007rw,Manuel:2007pz,Mannarelli:2008je,Mannarelli:2009ia,Dong:2007ax,Alford:2008pb,arXiv:0806.1005}.

Compact stars are strongly magnetized. Neutron star observations
indicate that the magnetic field is of the order of $B\sim
10^{12}-10^{13}$ G at the surface of ordinary pulsars. Magnetars -
strongly magnetized neutron stars - may feature even stronger
magnetic fields of the order of $10^{15}-10^{16}$ G
~\cite{Duncan:1992hi,Thompson:1993hn,Thompson:1995gw,Thompson:1996pe,Kouveliotou:1998fd,astro-ph/0011148,astro-ph/0111516}.
An upper limit on the magnetic field can be set through the virial
theorem. Gravitational equilibrium of stars is compatible with
magnetic fields of the order of
$10^{18}-10^{20}$~G~\cite{Astrophys:j383:745:1991,astro-ph/9703034,astro-ph/9703066}.
In such a strong magnetic field, not only the thermodynamical but
also the hydrodynamical properties of stellar matter will be
significantly affected. In particular, due to the large
magnetization of strange quark matter the fluid will be strongly
anisotropic in a strong magnetic field (we note here that the
magnetization of ordinary neutron matter is
small~\cite{astro-ph/0001537}). Therefore, there is need to develop
an anisotropic hydrodynamic theory to describe strongly magnetized
matter in compact stars. As we show below, the matter is completely
described in terms of eight viscosity coefficients, which include
six shear viscosities and two bulk viscosities.

In this paper, we will carry out a theoretical study of the
anisotropic hydrodynamics of magnetized strange quark matter and
will calculate the two bulk viscosities. We will also discuss the
implications of the anisotropic bulk viscosities on the r-mode
instability in rotating quark stars.

The paper is organized as follows. The formalism of anisotropic
hydrodynamics for magnetized strange quark matter is developed in
Sec.~\ref{2}. In Sec.~\ref{3} we apply the local linear response
method to derive explicit expressions for bulk viscosities. The
stability of the fluid under strong magnetic field is analyzed in
Sec.~\ref{4}. Section~\ref{5} contains our numerical results for the
bulk viscosities. The damping of the r-mode instability in rotating
quark stars by the bulk viscosity is studied in Sec.~\ref{6}.
Section~\ref{7} contains our summary. We use natural units
$\hbar=k_B=c=1$. The metric tensor is $g^{\m\n}=\diag(1,-1,-1,-1)$.
We will use the SI system of units in our equations involving
electromagnetism, however we will quote the strength of the magnetic
field in CGS units (Gauss), as is common in the literature on
compact stars.

%%%%%%%%%%%%%%%%%%%%%%%%%%%%%%%%%%%%%%%%%%%%%%%%%%%%%%%%%%%%%%%%%%%%%%%
\section {Anisotropic Hydrodynamics}\label{2}
%%%%%%%%%%%%%%%%%%%%%%%%%%%%%%%%%%%%%%%%%%%%%%%%%%%%%%%%%%%%%%%%%%%%%%%
%%%%%%%%%%%%%%%%%%%%%%%%%%%%%%%%%%%%%%%%%%%%%%%%%%%%%%%%%%%%%%%%%%%%%%%
\subsection {Ideal Hydrodynamics}\label{21}
%%%%%%%%%%%%%%%%%%%%%%%%%%%%%%%%%%%%%%%%%%%%%%%%%%%%%%%%%%%%%%%%%%%%%%%
Hydrodynamics arises as an effective theory valid in the
long-wavelength, low-frequency limit where the energy-momentum
tensor $T^{\m\n}$, the conserved baryon current $n_B^\m$, the
conserved electric current $n_e^\m$, and the entropy density flux
$s^\m$, {\it etc.}, are expanded in terms of gradients of the
4-velocity $u^\m$ and the thermodynamic parameters of the system,
such as the temperature $T$, baryon chemical potential $\m_B$, {\it
etc.}. The hydrodynamic equations can be expressed as conservation
laws for the total energy-momentum tensor $T^{\m\n}$, as well as
baryon and electric currents, $n_B^\m$ and $n_e^\m$. The zeroth
order terms in the expansion correspond to an ideal fluid and we
shall use the index 0 to label them. In the presence of an
electromagnetic field, the zeroth-order terms can be generally
written as~\cite{Groot:1969,Caldarelli:2008ze},
\begin{eqnarray}
\label{tmn00} T^{\m\n}_0&=&T_{\rm F0}^{\m\n}+T_{\rm EM}^{\m\n},\non
T^{\m\n}_{\rm F0}&=&\ve\, u^\m u^\n-P\D^{\m\n}-\frac{1}{2}\lb
M^{\m\l}F^{\;\;\n}_{\l}+M^{\n\l}F^{\;\;\m}_{\l}\rb,\non
n_{B0}^\m&=&n_B\, u^\m,\non n_{e0}^\m&=&n_e\, u^\m,\non s_0^\m &=&
s\, u^\m,
\end{eqnarray}
where $\ve$, $P$, $n_B$, $n_e$, and $s$ are the local energy
density, thermodynamic pressure, baryon number density, electric
charge density, and entropy density, respectively measured in the
rest frame of the fluid. $\D^{\m\n}\equiv g^{\m\n}-u^\m u^\n$ is the
projector on the directions orthogonal to $u^\m$.

Here $T_{\rm
EM}^{\m\n}=-F^{\m\l}F_{\;\;\l}^\n+g^{\m\n}F^{\r\s}F_{\r\s}/4$ is the
energy-momentum tensor of the electromagnetic field. $F^{\m\n}$ is
the field-strength tensor which can be decomposed into components
parallel and perpendicular to $u^\m$ as
\begin{eqnarray}
\label{fmn} F^{\m\n}&=&F^{\m\l}u_\l u^\n-F^{\n\l}u_\l
u^\m+\D^\m_{\;\;\a}F^{\a\b}\D_\b^{\;\;\n}\non &\equiv&E^\m u^\n-E^\n
u^\m+\frac{1}{2}\e^{\m\n\a\b}\lb u_\a B_\b-u_\b B_\a\rb,\non
\end{eqnarray}
where in the second line we have introduced the 4-vectors
$E^\m\equiv F^{\m\n}u_\n$ and $B^\m\equiv
\e^{\m\n\a\b}F_{\n\a}u_\b/2$ with $\e^{\m\n\a\b}$ being the totally
anti-symmetric Levi-Civita tensor. In the rest frame of the fluid,
$u^\m=(1,{\bf 0})$, we have $E^0=B^0=0$, $E^i=F^{i0}$ and
$B^i=-\e^{ijk}F_{jk}/2$, which are precisely the electric and
magnetic fields in this frame. Therefore, $E^\m$ and $B^\m$ are
nothing but the electric and magnetic fields measured in the frame
where the fluid moves with a velocity $u^\m$.

The antisymmetric tensor $M^{\m\n}$ is the polarization tensor which
describes the response to the applied field strength $F^{\m\n}$. For
example, if $\O$ is the thermodynamic potential of the system,
$M^{\m\n}\equiv -\pt\O/\pt F_{\m\n}$. For later use, we also define
the in-medium field strength tensor $H^{\m\n}\equiv
F^{\m\n}-M^{\m\n}$. In analogy to $F^{\m\n}$ we can decompose
$M^{\m\n}$ and $H^{\m\n}$ as
\begin{eqnarray}
M^{\m\n}&=&\lb P^\n u^\m-P^\m u^\n\rb+\frac{1}{2}\e^{\m\n\a\b}\lb
M_\b u_\a-M_\a u_\b\rb, \non H^{\m\n}&=&\lb D^\m u^\n-D^\n
u^\m\rb+\frac{1}{2}\e^{\m\n\a\b}\lb H_\b u_\a-H_\a u_\b\rb,\non
\end{eqnarray}
with $P^\m\equiv -M^{\m\n}u_\n$, $M^\m\equiv
\e^{\m\n\a\b}M_{\n\a}u_\b/2$, $D^\m\equiv H^{\m\n}u_\n$ and
$H^\m\equiv \e^{\m\n\a\b}H_{\n\a}u_\b/2$.

In the rest frame of the fluid, the non-trivial components of these
tensors are $(F^{10}, F^{20}, F^{30})=\bf{E}$, $(F^{32}, F^{13},
F^{21})=\bf{B}$, $(M^{10}, M^{20}, M^{30})=-\bf{P}$, $(M^{32},
M^{13}, M^{21})=\bf{M}$, $(H^{10}, H^{20}, H^{30})=\bf{D}$, and
$(H^{32}, H^{13}, H^{21})=\bf{H}$. Here $\bf{P}$ and $\bf{M}$ are
the electric polarization vector and magnetization vector,
respectively. In the linear approximation they are related to the
fields $\bf{E}$ and $\bf{B}$ by ${\bf P}=\c_e{\bf E}$ and ${\bf
M}=\c_m{\bf B}$, with $\c_e$ and $\c_m$ being the electric and magnetic
susceptibilities. The 4-vectors $E^\m, B^\m, \cdots$ are all space-like,
$E^\m u_\m=0, B^\m u_\m=0, \cdots$, and normalized as $E^\m
E_\m=-E^2, B^\m B_\m=-B^2, \cdots$, where $E\equiv |\bf E|$ and
$B\equiv |\bf B|$.

Since the electric field is much weaker than the magnetic field in
the interior of a neutron star, we will neglect it in most of the
following discussion. Upon introducing the 4-vector $b^\m\equiv
B^\m/B$, which is parallel to $B^\m$ and is normalized by the
condition $b^\m b_\m=-1$, and the antisymmetric tensor
$b^{\m\n}\equiv\e^{\m\n\a\b} b_\a u_\b$, we can write
\begin{eqnarray}
F^{\m\n}&=&-B b^{\m\n},\non M^{\m\n}&=&-M b^{\m\n},\non
H^{\m\n}&=&-H b^{\m\n},
\end{eqnarray}
with $M\equiv |\bf M|$ and $H\equiv |\bf H|$.

The Maxwell equation $\e^{\m\n\a\b}\pt_\b F_{\n\a}=0$ takes the form
\begin{eqnarray}
\label{maxwell0} \pt_\n\lb B^\m u^\n-B^\n u^\m\rb=0.
\end{eqnarray}
Its non-relativistic form, which is known as the induction
equation, is given by
\begin{eqnarray}
\frac{\pt \bf{B}}{\pt
t}&=&\nabla\times(\bv\times\bf{B}),\non\nabla\cdot{\bf B}&=&0,
\end{eqnarray}
where $\bv$ is the 3-velocity of the fluid. Contracting
\eq{maxwell0} with $b_\m$ gives
\begin{eqnarray}
\label{maxwell}\h+D\ln B-u^\n b^\m\pt_\m b_\n=0,
\end{eqnarray}
where $\h\equiv\pt_\m u^\m$ and $D\equiv u^\m\pt_\m$. The second
Maxwell equation can be written as
\begin{eqnarray}
\label{maxwell2}\pt_\m H^{\m\n}=n_e^\n,
\end{eqnarray}
whose non-relativistic form is
\begin{eqnarray}
\nabla\cdot\bf{D}-\bf{H}\cdot\nabla\times\bv&=&n_e^0,
\non\nabla\times\lb\bf{H}-\bf{D}\times\bv\rb-\frac{\pt\bf{D}}{\pt
t}-{\bf{H}}\times\pt_t\bv&=&{\bf{n}}_e,
\end{eqnarray}
where $n_{e}^{0}$ is the electric charge density and ${\bf n}_e$ is
the corresponding current. It is useful to re-write the
energy-momentum tensor in the following
form~\cite{Gedalin:1991,pre51:4901:1995,Sadooghi:2009fi},
\begin{eqnarray}
\label{tmn0} T^{\m\n}_{\rm F0}&=&\ve u^\m
u^\n-\pperp\Xi^{\m\n}+\ppara b^\m b^\n,\non T^{\m\n}_{\rm
EM}&=&\frac{1}{2}B^2\lb u^\m u^\n-\Xi^{\m\n}-b^\m
b^\n\rb,\non\Xi^{\m\n}&\equiv&\D^{\m\n}+b^\m b^\n,
\end{eqnarray}
where $\Xi^{\m\n}$ is the projection tensor on the direction
perpendicular to both $u^\m$ and $b^\m$. We have defined the
transverse and longitudinal pressures $\pperp=P-MB$ and $\ppara=P$
relative to $b^\m$; here $P$ is the thermodynamic pressure. In the
absence of a magnetic field, the fluid is isotropic and
$\pperp=\ppara=P$. In the local rest frame of fluid, we have
$b^\m=(0,0,0,1)$ (without loss of generality, we choose the $z$-axis
along the direction of the magnetic field), hence the
electromagnetic tensor takes the usual form, while $T^{\m\n}_{\rm
F0}=\diag(\ve,\pperp, \pperp, \ppara)$.

Next we would like to check the consistency of the terms that appear
in $T_{\rm F0}^{\m\n}$ with the formulae of standard thermodynamics
involving electromagnetic fields. By using the thermodynamic
relation
\begin{eqnarray}
\label{thermo} \ve&=&Ts+\m_B n_B+\m_e n_e-P,
\end{eqnarray}
and the conservation equations for $n_{B0}^\m$, $n_{e0}^\m$, and
$s_0^\m$ in ideal hydrodynamics, one can show that the hydrodynamic
equation $u_\n\pt_\m T^{\m\n}_0=0$ together with the Maxwell
equation (\ref{maxwell}) implies
\begin{eqnarray}
\label{de} D\ve=TDs+\m_B Dn_B+\m_e D n_e-MDB,
\end{eqnarray}
which is consistent with the standard thermodynamic relation
\begin{eqnarray}
\label{de2} d\ve&=&Tds+\m_B dn_B+\m_e dn_e-MdB.
\end{eqnarray}
One should note that the potential energy $-MB$ has already been
included in our definition of $\ve$. Otherwise, new terms $-MB$,
$-D(MB)$ and $-d(MB)$ should be added to the left-hand sides of
\eq{thermo}, \eq{de}, and \eq{de2}, respectively. Thus, we conclude
that our hydrodynamical equations are consistent with well-known
thermodynamic relations.

%%%%%%%%%%%%%%%%%%%%%%%%%%%%%%%%%%%%%%%%%%%%%%%%%%%%%%%%%%%%%%%%%%%%%%%
\subsection {Navier-Stokes-Fourier-Ohm Theory}\label{22}
%%%%%%%%%%%%%%%%%%%%%%%%%%%%%%%%%%%%%%%%%%%%%%%%%%%%%%%%%%%%%%%%%%%%%%%
By keeping the first-order terms of the derivative expansion of
conserved quantities one obtains the Navier-Stokes-Fourier-Ohm
theory. In this theory, $T^{\m\n}$, $n_B^\m$, $n_e^\m$, and $s^\m$
can be generally expressed as
\begin{eqnarray}
\label{tmn} T^{\m\n}&=&T_0^{\m\n}+h^\m u^\n+h^\n u^\m+\t^{\m\n},\non
n_B^\m&=&n_Bu^\m+j_B^\m,\non n_e^\m&=&n_eu^\m+j_e^\m,\non s^\m&=&s
u^\m+j_s^\m,
\end{eqnarray}
where $h^\m, \t^{\m\n}, j_B^\m, j_e^\m$, and $j_s^\m$ are the
dissipative fluxes. They all are orthogonal to $u^\m$; this reflects
the fact that the dissipation in the fluid should be spatial. We
shall assume that $j_s^\m$ can be expressed as a linear combination
of $h^\m, j^\m_B$, and $j^\m_e$~\cite{Hiscock:1985zz,Israel:1979wp}.
This allows us to incorporate the fact that the entropy flux is
determined by the energy-momentum and baryon number diffusion
fluxes. Thus,
\begin{eqnarray}
j_s^\m=\g h^\m-\a_B j^\m_B-\a_e j^\m_e,
\end{eqnarray}
with the coefficients $\g$, $\a_e$, and $\a_B$ being functions of
thermodynamic variables.

Next, the hydrodynamic equations are specified by utilizing the
conservation laws of the total energy-momentum $T^{\m\n}$, the
baryon number density flow $n_B^\m$,  electric current $n_e^\m$, and
the second law of thermodynamics,
\begin{eqnarray}
\label{hydroeqn} \pt_\m T^{\m\n}&=&0,\non \pt_\m n_B^\m&=&0,\non
\pt_\m n_e^\m&=&0,\non
 T\pt_\m s^\m&\geq&0.
\end{eqnarray}

To discuss the dissipative parts, let us first define the 4-velocity
$u^\m$, since it is not unique when energy exchange by thermal
conduction is allowed for. We will use the Landau-Lifshitz frame in
which $u^\m$ is chosen to be parallel to the energy density flow, so
that $h^\m=0$. Upon projecting the first equation of \eq{hydroeqn}
on $u^\n$ and after some straightforward manipulations, we find
\begin{eqnarray}
\label{energyeqn} (\ve+P)\h+D\ve-\t^{\m\n}\pt_\m u_\n+MDB=j_e^\l
u^\n F_{\n\l}.
\end{eqnarray}
Combining \eq{energyeqn}, \eq{thermo}, and the second equation in
\eq{hydroeqn}, we arrive at
\begin{eqnarray}
\label{divs} T\pt_\m s^\m&=&\t^{\m\n}w_{\m\n}+(\m_B-T\a_B)\pt_\m
j^\m_B-Tj^\m_B\nabla_\m\a_B\non&&\!\!\!\!\!\!+(\m_e-T\a_e)\pt_\m
j^\m_e-j^\m_e(T\nabla_\m\a_e+E_\m),
\end{eqnarray}
where $\nabla_\m\equiv \D_{\m\n}\pt^\n$ and $w^{\m\n}\equiv
\frac{1}{2}\lb\nabla^\m u^\n+\nabla^\n u^\m\rb$. For a
thermodynamically and hydrodynamically stable system, \eq{divs}
should be non-negative. This implies
\begin{eqnarray}
\a_B&=&\b\m_B,\non \a_e&=&\b\m_e,\non
\t^{\m\n}&=&\w^{\m\n\a\b}w_{\a\b},\non
j_B^\m&=&-\k^{\m\n}T\nabla_\n\a_B,\non
j_e^\m&=&-\s^{\m\n}(T\nabla_\n\a_e+E_\n),
\end{eqnarray}
where $\b\equiv1/T$, $\w^{\m\n\a\b}$ is the rank-four tensor of
viscosity coefficients, and $\k^{\m\n}$ and $\s^{\m\n}$ are thermal
and electrical conductivity tensors with respect to the diffusion
fluxes of baryon number density and electric charge density. By
definition, $\w^{\m\n\a\b}$ is symmetric in the pairs of indices
$\a,\b$ and $\m,\n$. It necessarily satisfies the condition
$\w^{\m\n\a\b}(B^\s)=\w^{\a\b\m\n}(-B^\s)$, which is Onsager's
symmetry principle for transport coefficients. Similarly, the
tensors $\k^{\m\n}$ and $\s^{\m\n}$ should satisfy the conditions
$\k^{\m\n}(B^\l)=\k^{\n\m}(-B^\l)$ and
$\s^{\m\n}(B^\l)=\s^{\n\m}(-B^\l)$. Furthermore, all the tensors of
transport coefficients $\w^{\m\n\a\b}$, $\k^{\m\n}$, and $\s^{\m\n}$
must be orthogonal to $u^\m$ by definition.

As we have seen, the appearance of the magnetic field makes the
system anisotropic. Such anisotropy is specified by the vector
$b^\m$, so that the tensors $\w^{\m\n\a\b}$, $\s^{\m\n}$, and
$\k^{\m\n}$ should be in general expressed in terms of $u^\m$,
$b^\m$, $g^{\m\n}$, and $b^{\m\n}$. All independent irreducible
tensor combinations having the symmetry of $\w^{\m\n\a\b}$ and which
are orthogonal to $u^\m$ are~\cite{Lifshitz:1979}(see also Appendix
\ref{proof})
\begin{eqnarray}
\label{combination}
&&({\rm{i}})\quad \D^{\m\n}\D^{\a\b},\non
&&({\rm{ii}})\quad \D^{\m\a}\D^{\n\b}+\D^{\m\b}\D^{\n\a},\non
&&({\rm{iii}})\quad\D^{\m\n}b^\a b^\b+\D^{\a\b}b^\m b^\n,\non
&&({\rm{iv}})\quad b^\m b^\n b^\a b^\b,\non
&&({\rm{v}})\quad \D^{\m\a}b^\n
b^\b+\D^{\n\b}b^\m b^\a+\D^{\m\b}b^\n b^\a+\D^{\n\a}b^\m b^\b,\non
&&({\rm{vi}})\quad
\D^{\m\a} b^{\n\b}+\D^{\n\b} b^{\m\a}+\D^{\m\b} b^{\n\a}
+\D^{\n\a} b^{\m\b},\non
&&({\rm{vii}})\quad b^{\m\a}b^\n b^\b
+b^{\n\b}b^\m b^\a+b^{\m\b}b^\n b^\a+b^{\n\a}b^\m b^\b,\non
&&({\rm{viii}})\quad b^{\m\a}b^{\n\b}+b^{\m\b}b^{\n\a}.
\end{eqnarray}
All independent irreducible tensor combinations having the symmetry
of $\k^{\m\n}$ and $\s^{\m\n}$ and which are orthogonal to $u^\m$
are
\begin{eqnarray}
\label{combination-k} &&({\rm{i}})\;\; \D^{\m\n},\non
&&({\rm{ii}})\;\; b^\m b^\n,\non
&&({\rm{iii}})\;\;b^{\m\n}.
\end{eqnarray}
In accordance with the number of tensors (\ref{combination}) and
(\ref{combination-k}), a fluid in a magnetic field in general has
eight independent viscosity coefficients, three independent thermal
conduction coefficients and three independent electrical
conductivities. They may be defined as the coefficients in the
following decompositions for the viscous stress tensor, heat flux,
and electric charge flux
\begin{widetext}
\begin{eqnarray}
\label{pmn}\t^{\m\n}&=&2\w_0\lb
w^{\m\n}-\D^{\m\n}\h/3\rb+\w_1\lb\D^{\m\n}-\frac{3}{2}\Xi^{\m\n}\rb\lb\h-\frac{3}{2}\f\rb-2\w_2\lb
b^\m\Xi^{\n\a}b^\b+b^\n\Xi^{\m\a}b^\b\rb w_{\a\b}\non &-&\w_3\lb
2b^{\m\a}b^{\n\b}w_{\a\b}-\Xi^{\m\a}w_\a^{\;\;\n}-\Xi^{\n\a}w_\a^{\;\;\m}\rb-2\w_4\lb
\Xi^{\m\a}b^{\n\b}+\Xi^{\n\a}b^{\m\b}\rb w_{\a\b}\non &+&2\w_5\lb
b^{\m\a}b^\n b^\b+b^{\n\a}b^\m b^\b\rb w_{\a\b}
+\frac{3}{2}\zperp\Xi^{\m\n}\f+3\zpara b^\m b^\n \vf,
\end{eqnarray}
\begin{eqnarray}
\label{jbm} j_B^\m&=&\k T\nabla^\m\a_B-\k_1 b^\m b^\n T\nabla_\n\a_B
-\k_2 b^{\m\n} T\nabla_\n\a_B,
\end{eqnarray}
\begin{eqnarray} \label{jbe} j_e^\m&=&\s
(\nabla^\m\a_e+E^\m)-\s_1 b^\m b^\n (\nabla^\m\a_e+E^\m) -\s_2
b^{\m\n} (\nabla^\m\a_e+E^\m),
\end{eqnarray}
\end{widetext}
where $\f\equiv \Xi^{\m\n}w_{\m\n}$, $\vf\equiv b^\m b^\n w_{\m\n}$
and $\t^{\m\n}$ is constructed so that the $\w$'s are the
coefficients of its traceless parts, \ie, they can be regarded as
shear viscosities; $\z$'s are the coefficients of the parts with
non-zero trace and can be considered as bulk viscosities. The $\k$'s
and $\s$'s are thermal and electrical conductivities, respectively.

Now the divergence of entropy density flux (\ref{divs}) can be
explicitly written as
\begin{widetext}
\begin{eqnarray}
\label{divs2} T\pt_\m s^\m&&\!\!\!\!\!\!=2\w_0\lb
w^{\m\n}-\frac{1}{3}\D^{\m\n}\h\rb\lb
w_{\m\n}-\frac{1}{3}\D_{\m\n}\h\rb+\w_1\lb\h-\frac{3}{2}\f\rb^2\non
&+&2\w_2\lb b^\m b_\r w^{\r\n}-b^\n b_\r w^{\r\m}\rb\lb b_\m b^\r
w_{\r\n}-b_\n b^\r w_{\r\m}\rb\non &+&\w_3\lb
b^{\m\r}w_\r^{\;\;\n}-b^{\n\r}w_\r^{\;\;\m}\rb\lb
b_{\m\r}w^\r_{\;\;\n}-b_{\n\r}w^\r_{\;\;\m}\rb\non
&+&\frac{3}{2}\zperp\f^2+3\zpara\vf^2-\k
T^2\nabla^\m\a_B\nabla_\m\a_B+\k_1 T^2\lb b^\m\nabla_\m\a_B\rb^2\non
&-&\s(T\nabla^\m\a_e+E^\m)(T\nabla_\m\a_e+E_\m)+\s_1(Tb^\m\nabla_\m\a_e+E^\m
b_\m)^2.
\end{eqnarray}
\end{widetext}
One should note that the terms corresponding to the transport
coefficients $\w_4,\w_5$, $\k_2$, and $\s_2$ in \eqs{pmn}{jbe} do
not contribute to the divergence of the entropy density flux. For
stable systems, all the other transport coefficients must be
positive definite according to the second law of thermodynamics. In
Sec.\ref{4} we will demonstrate explicitly that negative bulk
viscosities $\zperp$ or/and $\zpara$ indeed cause an instability in
the hydrodynamic evolution of strange stars.

To conclude this section, we compare our definition of the viscosity
coefficients in \eq{pmn} with the definition given in
Ref.~\cite{Lifshitz:1979} for non-relativistic fluid, which reads
\begin{widetext}
\begin{eqnarray}
\label{pij}\t_{ij}&=&2\tilde{\w}\lb
w_{ij}-\d_{ij}\h/3\rb+\tilde{\z}\d_{ij}\h\non
&+&\tilde{\w}_1(2w_{ij}-\d_{ij}\h+\d_{ij}w_{kl}b_kb_l-2w_{ik}b_kb_j-2w_{jk}b_kb_i+b_ib_j\h+b_ib_jw_{kl}b_kb_l)\non
&+&2\tilde{\w}_2(w_{ik}b_kb_j+w_{jk}b_kb_i-2b_ib_jw_{kl}b_kb_l)+\tilde{\w}_3(w_{ik}b_{jk}+w_{jk}b_{ik}-w_{kl}b_{ik}b_jb_l-w_{kl}b_{jk}b_ib_l)\non
&+&2\tilde{\w}_4(w_{kl}b_{il}b_jb_k+w_{kl}b_{jl}b_ib_k)+\tilde{\z}_1(\d_{ij}w_{kl}b_kb_l+b_ib_j\h),
\end{eqnarray}
where $b_{ij}\equiv\e_{ijk}b_k$ and the remaining notations are self-explanatory.

Our viscosity coefficients in \eq{pmn} are related to the coefficients in \eq{pij} by
\begin{eqnarray}
\label{relate} &\w_0=\tilde{\w}+\tilde{\w}_1,\;\;
\w_1=\frac{3}{4}\lb\tilde{\w}_1+\frac{1}{2}\tilde{\z}_1-\frac{3}{2}\tilde{\z}\rb,\;\;\w_2=\tilde{\w}_2-\tilde{\w}_1,\;\;\w_4=\frac{1}{2}\tilde{\w}_3,&\non
&\w_5=\tilde{\w}_4,\;\;\zperp=\tilde{\z}+\frac{1}{3}\tilde{\z}_1,\;\;\zpara=\tilde{\z}+\frac{4}{3}\tilde{\z}_1.&
\end{eqnarray}
In Ref.~\cite{Lifshitz:1979} there is no term that corresponds
to our $\w_3$. The reason is that Ref.~\cite{Lifshitz:1979}
considers the combination of vectors (viii) in \eq{combination}
as dependent on the others; in
the Appendix we will show that (at least for relativistic fluids)
all the combinations (i)-(viii) are linearly independent.
Note that the transport coefficients in
Eq.~(\ref{divs2}) appear as prefactors of quadratic forms, therefore
the second law of thermodynamics requires that these coefficients
must be positive definite for stable ensembles. This is not manifest
in Eq.~(\ref{pij}).
\end{widetext}

%%%%%%%%%%%%%%%%%%%%%%%%%%%%%%%%%%%%%%%%%%%%%%%%%%%%%%%%%%%%%%%%%%%%%%%
\section {Bulk Viscosities}\label{3}
%%%%%%%%%%%%%%%%%%%%%%%%%%%%%%%%%%%%%%%%%%%%%%%%%%%%%%%%%%%%%%%%%%%%%%%
The typical oscillation frequency of neutron stars is of the order
of magnitude of the rotation frequency, $1
$s$^{-1}\lesssim\o\lesssim10^3$s$^{-1}$. The most important
microscopic processes which dissipate energy on the corresponding
timescales are the weak processes.

The compression and expansion of strange quark matter with nonzero
strange quark mass will drive the system out of equilibrium. The
processes (\ref{weak1}) and (\ref{weak2}) are the most efficient
microscopic processes that restore local chemical equilibrium.
Therefore, the bulk viscosities are determined mainly by the
processes (\ref{weak1}) and (\ref{weak2}). In this section we will
derive analytical expressions for the bulk viscosities $\zperp$ and
$\zpara$
%--------------------------- footnote
\footnote{All the other transport coefficients of strange quark
matter are dominated by strong processes. In leading order in the
QCD coupling constant $\a_s\equiv g^2/4\p$, these processes involve
quark-quark scattering via an interchange of a single gluon
~\cite{Heiselberg:1993cr}. However, these processes are too fast to
contribute to the bulk viscosity at the frequency of the order of
the stellar oscillation frequency.}.

Let us imagine an isotropic flow $\bv(t)\sim e^{i\o t}$ which
characterizes the stellar oscillation. If there are no dissipative
processes, such an oscillation will drive the system from one
instantaneous equilibrium state to another instantaneous equilibrium
state. The appearance of dissipation changes the picture: during the
oscillations the thermodynamic quantities will differ from their
equilibrium values. Let us explore how the thermodynamic quantities
evolve during the flow oscillation.

In  general, we can write the change of baryon density
$n_B\equiv(n_u+n_d+n_s)/3$ induced by the oscillation of the fluid
as
\begin{eqnarray}
n_B(t)&=&n_{B0}+\d n_{B}(t),\non \d n_{B}&=&\d n_B^{\rm eq}+\d
n'_{B},
\end{eqnarray}
where $n_{B0}$ is the static (time-independent) equilibrium value,
$\d n_B^{\rm eq}$ denotes the equilibrium value shift from $n_{B0}$
due to the volume change and $\d n'_B$ denotes the instantaneous
departure from the equilibrium value. Because processes
(\ref{weak1}) and (\ref{weak2}) conserve baryon number, $\d
n'_{B}(t)$ can be set to zero,  if  we neglect other microscopic
processes. Then $\d n_B$ can be determined through the continuity
equation of ideal hydrodynamics,
\begin{eqnarray}
\label{deltanb} \d n_B(t)&=&-\frac{n_{B0}}{i\o}\h.
\end{eqnarray}
Since processes (\ref{weak1}) and (\ref{weak2}) also conserve the
sum $n_d+n_s$, a similar argument leads to the relation
\begin{eqnarray}
\label{number} \d n_d+\d n_s&=&-\frac{n_{d0}+n_{s0}}{i\o}\h.
\end{eqnarray}

When the system is driven out of chemical equilibrium, the chemical
potential of the $s$ quark will be slightly different from that of
the $d$ quark. Let us denote this difference by
$\d\m=\m_s-\m_d=\d\m_s-\d\m_d$, with $\d\m_f$ being the deviation of
$\m_f$ from its static equilibrium value. Up to linear order in the
deviation we find
\begin{eqnarray}
\label{deltamu}\d \m(t)\simeq\lb\frac{\pt\m_s}{\pt n_s}\rb_0\d
n_s-\lb\frac{\pt\m_d}{\pt n_d}\rb_0\d n_d,
\end{eqnarray}
where $n_f$ denotes the number density of quarks of flavor $f$ and
the subscript 0 indicates that the quantity in the bracket is
computed in static equilibrium state. $\d n_s$ and $\d n_d$ are the
deviations of $s$-quark and $d$-quark densities from their static
equilibrium value. In the final expressions they should be functions
of $\h$.

The instantaneous departure from equilibrium is restored by
the weak processes (\ref{weak1}) and (\ref{weak2}). Adopting the
linear approximation, this can be described by
\begin{eqnarray}
\label{rate} \G_d-\G_s=\l\d\m, \;\;\;\;\l>0,
\end{eqnarray}
where $\G_d$ and $\G_s$ are the rates of processes (\ref{weak1}) and
(\ref{weak2}), respectively. If the weak processes are turned off,
one should have
\begin{eqnarray}
\dot{\d n_f^{\rm eq}}=-n_{f0}\h=n_{f0}\frac{\dot{\d n_B}}{n_{B0}},
\end{eqnarray}
where the dot denotes the time derivative. After turning on the weak
processes, we have
\begin{eqnarray}
\label{dtn} \dot{\d n_u}&=&\dot{\d n_B},\non \dot{\d
n_d}&=&n_{d0}\frac{\dot{\d n_B}}{n_{B0}}+\l\d\m(t),\non \dot{\d
n_s}&=&n_{s0}\frac{\dot{\d n_B}}{n_{B0}}-\l\d\m(t).
\end{eqnarray}
This system of coupled linear first-order equations is closed by
substituting \eqs{deltanb}{deltamu}. It is then easy to obtain the
solution,
\begin{eqnarray}
\label{deltand} \d n_u&=& -n_{u0}\frac{\h}{i\o},\non \d
n_{d}&=&-\frac{i\o n_{d0}+\l\lb\pt\m_s/\pt n_s\rb_0(
n_{d0}+n_{s0})}{i\o+\l A}\frac{\h}{i\o},\non \d n_{s}&=&-\frac{i\o
n_{s0}+\l\lb\pt\m_d/\pt n_d\rb_0( n_{d0}+n_{s0})}{i\o+\l
A}\frac{\h}{i\o},\non
\end{eqnarray}
where the coefficient $A$ is defined by
\begin{eqnarray}
\label{A} A=\lb\frac{\pt\m_s}{\pt n_s}\rb_0+\lb\frac{\pt\m_d}{\pt
n_d}\rb_0.
\end{eqnarray}

The parallel and transverse components of the pressure $\ppara$ and
$\pperp$ can be written as
\begin{eqnarray}
\ppara &=& \ppara^{\rm eq}+\d \ppara',\non \pperp &=& \pperp^{\rm
eq}+\d \pperp',
\end{eqnarray}
and
\begin{eqnarray}
\ppara_{/\perp}^{\rm eq} &\simeq&
\ppara_{/\perp}^0+\sum_f\lb\frac{\pt\ppara_{/\perp}}{\pt n_f}\rb_0\d
n_{f}^{\rm eq}+\lb\frac{\pt\ppara_{/\perp}}{\pt B}\rb_0\d B,\non \d
\ppara_{/\perp}'&\simeq&\sum_f\lb\frac{\pt\ppara_{/\perp}}{\pt
n_f}\rb_0\d n_f',
\end{eqnarray}
where
\begin{eqnarray}
\d n_f'&\equiv&\d n_f-\d n_f^{\rm eq}.
\end{eqnarray}
The small departure of the magnetic field $\d B$ can be calculated
by the variation of \eq{maxwell}. One finds
\begin{eqnarray}
\d B=-\frac{2}{3}\frac{B}{i\o}\h.
\end{eqnarray}
A direct calculation then gives,
\begin{eqnarray}
\d n_u'&=&0,\non \d n_d'&=&\frac{\l\cpara}{i\o+\l
A}\frac{\h}{i\o},\non \d n_s'&=-&\frac{\l\cpara}{i\o+\l
A}\frac{\h}{i\o},
\end{eqnarray}
where we introduced the coefficient $\cpara$ as
\begin{eqnarray}
\cpara&\simeq&n_{d0}\lb\frac{\pt\m_d}{\pt
n_d}\rb_0-n_{s0}\lb\frac{\pt\m_s}{\pt n_s}\rb_0.
\end{eqnarray}
Now we obtain
\begin{eqnarray}
\d \ppara_{/\perp}'&=&-\frac{\l\cpara\cpara_{/\perp}}{i\o+\l
A}\frac{\h}{i\o},
\end{eqnarray}
with $\cperp$ defined as
\begin{eqnarray}
\cperp&\simeq&\cpara-XB,\non X&=&\lb\frac{\pt\cm}{\pt
n_d}\rb_0-\lb\frac{\pt\cm}{\pt n_s}\rb_0.
\end{eqnarray}

Then the deviation of $T^{\m\n}$ from its equilibrium value can be
written as
\begin{eqnarray}
\d T^{\m\n}=-\re\d \pperp'\Xi^{\m\n}+\re\d \ppara' b^\m b^\n.
\end{eqnarray}
For  isotropic flows,  we have
\begin{eqnarray}
\t^{\m\n}=\zperp\Xi^{\m\n}\h-\zpara b^\m b^\n\h.
\end{eqnarray}
By comparing the above two expressions, we obtain
\begin{eqnarray}
\label{vis1}\zpara=\frac{\l\cpara^2}{\o^2+\l^2A^2},
\end{eqnarray}
and
\begin{eqnarray}
\label{vis2}\zperp=\frac{\l\cperp\cpara}{\o^2+\l^2A^2}.
\end{eqnarray}
Expressions (\ref{vis1}) and (\ref{vis2}) show that the bulk
viscosities $\zperp$ and $\zpara$ are functions of the perturbation
frequency $\o$, the weak rate $\l$ and the thermodynamic quantities
$\cpara, \cperp, A$. From the derivation above we can convince
ourselves that these expressions should be valid also in the case of
color-superconducting matter. For zero magnetic field, \eq{vis2}
reduces to \eq{vis1} which, with parameters $\cpara, A$, and $\l$
taken in the absence of  magnetic field gives the expression for the
usual bulk viscosity $\z_0$ defined in isotropic hydrodynamics.

Both $\zperp$ and $\zpara$ attain their maxima in the limit of zero
frequency, $\zpara^{\rm max}=\cpara^2/(\l A^2),\;\zperp^{\rm
max}=\cpara\cperp/(\l A^2)$; and the maxima are inversely
proportional to the weak interaction rate.  At high frequency,
$\o\gg\l A$, $\zperp$ and $\zpara$ fall off as $1/\o^2$. For
practical applications to cold strange stars, where the chemical
potential is much larger than the temperature, the quantities
$\cpara, \cperp$, and $A$ can be evaluated in the zero-temperature
limit. Their dependence on temperature is weak. Contrary to this,
the coefficient $\l$ depends strongly on temperature: for normal
quark matter, $\l$ has a power-law dependence on $T$ (see
Sec.\ref{5}); for a fully paired color-superconducting phase, the
weak rate is exponentially suppressed by a Boltzmann factor
$e^{-\D/T}$ with $\D$ being the superconducting gap. Consequently,
$\zperp$ and $\zpara$ depend exponentially on
$T$~\cite{Sad:2006qv,Alford:2006gy,Sad:2007ud,Dong:2007mb,Alford:2007rw,Manuel:2007pz,Dong:2007ax,Alford:2008pb,arXiv:0806.1005}.

Before we find the numerical values for the bulk viscosities, we
first need to analyze the stability of magnetized strange quark
matter. We observe that according to \eq{vis2} negative values of
$\zperp$ are a priori not excluded. In the following we analyze the
consequences and implications of negative $\zperp$ on the stability
of the system.

%%%%%%%%%%%%%%%%%%%%%%%%%%%%%%%%%%%%%%%%%%%%%%%%%%%%%%%%%%%%%%%%%%%%%%%
\section {Stability Analysis}\label{4}
%%%%%%%%%%%%%%%%%%%%%%%%%%%%%%%%%%%%%%%%%%%%%%%%%%%%%%%%%%%%%%%%%%%%%%%
%%%%%%%%%%%%%%%%%%%%%%%%%%%%%%%%%%%%%%%%%%%%%%%%%%%%%%%%%%%%%%%%%%%%%%%
\subsection {Mechanical Stability}\label{41}
\begin{figure}[!htb]
\begin{center}
\includegraphics[width=7.5cm]{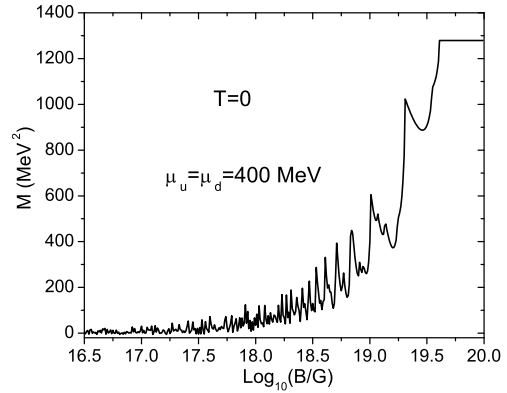}
\caption{The magnetization of strange quark matter as function of
the magnetic field $B$.} \label{mag_u}
\end{center}
\end{figure}

%%%%%%%%%%%%%%%%%%%%%%%%%%%%%%%%%%%%%%%%%%%%%%%%%%%%%%%%%%%%%%%%%%%%%%%
Stable equilibrium in a self-gravitating fluid, such as in strange
quark stars, is attained through the balance of gravity and
pressure. The gravitational equilibrium requires that both
components of the pressure $\ppara$ and $\pperp$ should be positive
(otherwise the star will undergo a gravitational collapse). At zero
temperature, the one-loop thermodynamic pressure $P=T{\ln \cal Z}/V$
of non-interacting strange quark matter, where $\cal Z$ is the grand
partition function, is given by
\begin{eqnarray}
\label{pstrong}
P&=&\sum_{f=u,d,s}\frac{N_cq_fB}{4\p^2}\sum_{n=0}^{n_{\rm
max}^f}\n_n
\Bigg[\m_f\sqrt{\m_f^2-m_f^2-2nq_fB}\non&&-(m_f^2+2q_fBn)\ln\frac{\m_f+\sqrt{\m_f^2-m_f^2-2q_fBn}}{\sqrt{m_f^2+2q_fBn}}\Bigg],\non
\end{eqnarray}
where $q_f$ is the absolute value of electric charge, $\m_f$ and
$m_f$ are the chemical potential and the mass of quark of flavor
$f$, $n$ labels the Landau levels, $\n_n=2-\d_{0n}$ is the degree of
degeneracy of each Landau level and $n_{\rm max}^f={\rm
Int}[(\m_f^2-m_f^2)/(2q_fB)]$ is the highest Landau level for quarks
of flavor $f$. By differentiating \eq{pstrong} with respect to $B$
one can easily get the magnetization as
\begin{eqnarray}
\label{M0} M&=&\sum_f\sum_{n=0}^{n_{\rm max}^f}\n_n\frac{N_c
q_f}{4\p^2}
\Bigg[\m_f\sqrt{\m_f^2-m_f^2-2nq_fB}\non&&\!\!\!\!\!-(m_f^2+4n q_f
B)\ln\frac{\m_f+\sqrt{\m_f^2-m_f^2-2n q_f B}}{\sqrt{m_f^2+2n q_f
B}}\Bigg].\non
\end{eqnarray}
In \fig{mag_u} we illustrate the magnetization as a function of
magnetic field at zero temperature. The parameters are chosen as
\begin{eqnarray}
\label{paras}
&& \m_u=\m_d=\m_s=400~{\rm MeV},\non
&& m_s=150~{\rm MeV},\non
&& m_u=m_d=5~{\rm MeV}.
\end{eqnarray}
On average, the magnetization increases when $B$ grows
and eventually becomes constant when $B>B_c$, where
\begin{eqnarray}
B_c\equiv{\rm Max}_f\{(\m_f^2-m_f^2)/(2 q_f)\}.
\end{eqnarray}
However, the detailed structure of the magnetization exhibits strong
de Haas-van Alphen oscillations~\cite{landau:statistical}. This
oscillatory behavior is of the same origin as the de Haas-van Alphen
oscillations of the magnetization in metals and originates from the
quantization of the energy levels associated with the orbital motion
of charged particles in a magnetic field. The irregularity of this
oscillation shown in \fig{mag_u} is due to the unequal masses and
charges of $u,d$, and $s$ quarks.

\begin{figure}[!htb]
\begin{center}
\includegraphics[width=7.5cm]{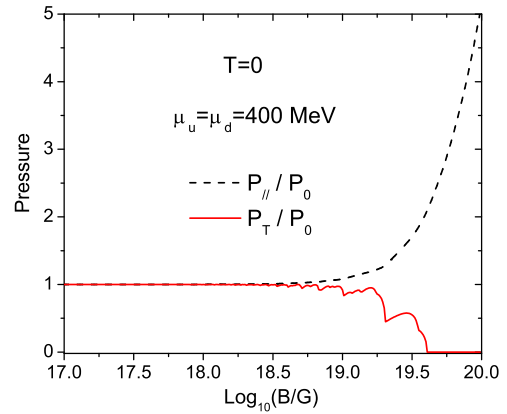}
\caption{(Color online) The  parallel $\ppara$ (dashed, black
online) and transverse $\pperp$ (solid, red online)  pressures of
strange quark matter as functions of the magnetic field $B$ in units
of the pressure $P_0$ for zero magnetic field.} \label{pressure_u}
\end{center}
\end{figure}

When $B>B_c$, all quarks are confined to their lowest Landau level
and their transverse motions are frozen. In this case, the
longitudinal pressure $\ppara\propto B$, so the magnetization
$M\equiv\pt P/\pt B=P/B$ is independent of $B$, and the transverse
pressure $\pperp=P-MB$ of the system vanishes. This behavior is
evident in \fig{pressure_u}.
%%%%%%%%%%%%%%%%%%%%%%%%%%%%%%%%%%%%%%%%%%%%%%%%%%%%%%%%%%%%%%%%%%%%%%%
%%%%%%%%%%%%%%%%%%%%%%%%%%%%%%%%%%%%%%%%%%%%%%%%%%%%%%%%%%%%%%%%%%%%%%%
%%%%%%%%%%%%%%%%%%%%%%%%%%%%%%%%%%%%%%%%%%%%%%%%%%%%%%%%%%%%%%%%%%%%%%%
%%%%%%%%%%%%%%%%%%%%%%%%%%%%%%%%%%%%%%%%%%%%%%%%%%%%%%%%%%%%%%%%%%%%%%%

Thus, we conclude that when all the quarks are confined to their
lowest Landau level, the transverse pressure vanishes. The system
therefore becomes mechanically unstable and would collapse due to
the gravity~\cite{hep-ph/9911218}. This phenomenon establishes an
upper limit on the magnetic field sustained by a quark star. Given
$\m_d\sim 400 $ MeV, this upper limit is roughly $B_c\sim10^{20}$ G
as shown in \fig{pressure_u}.

%%%%%%%%%%%%%%%%%%%%%%%%%%%%%%%%%%%%%%%%%%%%%%%%%%%%%%%%%%%%%%%%%%%%%%%
\subsection {Thermodynamic Stability}\label{42}
\begin{figure}[!htb]
\begin{center}
\includegraphics[width=7.5cm]{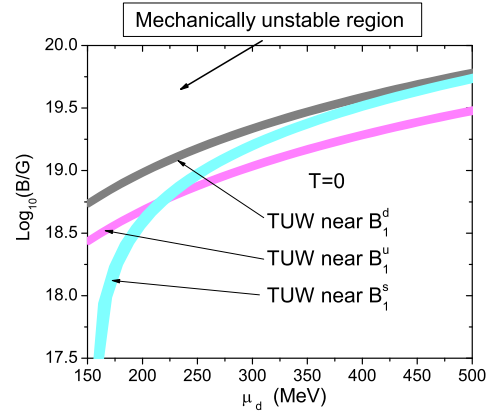}
\caption{(Color online) The first three thermodynamically unstable
windows (TUW) for strange quark matter in the $\log B-\m_d$ plane
for $T=0$ and $\mu_u=\mu_d$. } \label{TIW}
\end{center}
\end{figure}

%%%%%%%%%%%%%%%%%%%%%%%%%%%%%%%%%%%%%%%%%%%%%%%%%%%%%%%%%%%%%%%%%%%%%%%
The thermodynamic stability requires that the local entropy density
should reach its maximum in the equilibrium state~\cite{prigogine}. The
total energy density of the fluid and the magnetic field is
\begin{eqnarray}
\label{totaleps} \ve_{\rm total}=Ts+\m_f n_f-P+\frac{B^2}{2},
\end{eqnarray}
and the corresponding first law of thermodynamics, in variational
form, is
\begin{eqnarray}
\label{dtve} \d\ve_{\rm total}=T\d s+\m_f \d n_f+H\d B,
\end{eqnarray}
where $H$ is the strength of the magnetic field and $\d $ stands for
a small departure of a given quantity from its equilibrium value.
Varying \eq{dtve} on both sides and taking into account that
$\ve_{\rm total}, n_f$, and $B$ are independent variational
variables, one obtains
\begin{eqnarray}
\d^2s&=&-\frac{1}{T}\d\m_f\d n_f-\frac{1}{T}\d H\d B=-\frac{1}{T} \d
x^T \chi \d x,
\end{eqnarray}
where $\d x=\lb\d n_u,\;\d n_d,\; \d n_s,\;\d B\rb$ and
\begin{eqnarray}
\label{thermostab}\chi&=&\lb\begin{array}{cccc}\frac{\pt \m_u}{\pt
n_u} & 0 & 0 & \frac{\pt \m_u}{\pt B}\\0 & \frac{\pt \m_d}{\pt n_d}
& 0 & \frac{\pt \m_d}{\pt B}\\ 0& 0 & \frac{\pt \m_s}{\pt n_s} &
\frac{\pt \m_s}{\pt B} \\ \frac{\pt H}{\pt n_u} & \frac{\pt H}{\pt
n_d} & \frac{\pt H}{\pt n_s} & \frac{\pt H}{\pt B}\end{array}\rb_0.
\end{eqnarray}
The thermodynamical stability criteria require that $\d s=0,\;\d^2
s\leq 0$, or, equivalently, $\chi$ is positive definite. Taking into
account the relation $(\pt H/\pt n_f)_0=-(\pt \m_f/\pt B)_0$, it is
easy to show that these criteria are equivalent to the requirement
\begin{eqnarray}
\lb\frac{\pt n_f}{\pt \m_f}\rb_0\geq0,\;\;\lb\frac{\pt M}{\pt
B}\rb_0\leq 1.
\end{eqnarray}
From \eq{pstrong} we obtain
\begin{eqnarray}
\label{nf0} n_{f0}&=&\frac{N_cq_f B}{2\p^2}\sum_{n=0}^{n_{\rm
max}^f}\n_n\sqrt{\m_f^2-m^2_f-2nq_fB},\non \frac{\pt n_{f0}}{\pt
\m_f}&=&\frac{N_cq_fB}{2\p^2}\sum_{n=0}^{n_{\rm
max}^f}\n_n\frac{\m_f}{\sqrt{\m_f^2-m^2_f-2nq_fB}};
\end{eqnarray}
then it is evident that the condition $\lb\pt n_f/\pt
\m_f\rb_0\geq0$ is always satisfied. However, $\lb \pt M/\pt B\rb_0$
is divergent when $B$ approaches a $n\neq0$ Landau level for each
flavor quark from below,
\begin{eqnarray}
\lb\frac{\pt M}{\pt B}\rb_0&\ra& \frac{N_c q_f B}{\p^2}\frac{(n
q_f)^2}{\m_f\sqrt{\m_f^2-m_f^2-2n q_f B}},\non {\rm when}&& B\ra
B_n^f\equiv\frac{\m_f^2-m_f^2}{2 n q_f}-0^+.
\end{eqnarray}
This shows that strange quark matter will be thermodynamically
unstable just below each Landau level $B_n^f$ ($n\neq0$) for every
flavor $f$. The first three thermodynamically unstable windows (TUW)
associated, respectively, with $B^d_1, B^u_1$, and $B_1^s$ are
illustrated in the $\log B-\m$ plane in \fig{TIW} for our parameters
(\ref{paras}). The TUW is actually very narrow. One may conjecture
that such an instability may lead to formation  of magnetic
domains~\cite{landau:statistical}. The presence of possible magnetic
domains in neutron star crusts was discussed in
Ref.~\cite{jpc:14:6233:1982}; furthermore, such a possibility for
color-flavor-locked quark matter  was pointed out in
Ref.~\cite{arXiv:0708.0307}. We will not pursue here the study of
domain structure and related physics, since among other things, this
will require us to specify the  geometry of the system.
%%%%%%%%%%%%%%%%%%%%%%%%%%%%%%%%%%%%%%%%%%%%%%%%%%%%%%%%%%%%%%%%%%%%%%%
%%%%%%%%%%%%%%%%%%%%%%%%%%%%%%%%%%%%%%%%%%%%%%%%%%%%%%%%%%%%%%%%%%%%%%%

%%%%%%%%%%%%%%%%%%%%%%%%%%%%%%%%%%%%%%%%%%%%%%%%%%%%%%%%%%%%%%%%%%%%%%%
\subsection {Hydrodynamic Stability}\label{43}
%%%%%%%%%%%%%%%%%%%%%%%%%%%%%%%%%%%%%%%%%%%%%%%%%%%%%%%%%%%%%%%%%%%%%%%
In this subsection we address the problem of hydrodynamic stability within
the theory presented in Sec.\ref{2}.  The fluid is said to be stable if it
returns to its initial state after a transient perturbation. Otherwise,
\ie,  when the perturbation grows and takes the fluid into another state,
the fluid is unstable. Our particular goal here is to determine
whether a small,  plane-wave perturbation  around a
homogeneous equilibrium state grows for nonzero $\zperp$ and
$\zpara$~\cite{Hiscock:1985zz}. All the other transport coefficients
are set to zero. For this purpose, it is sufficient to solve the
hydrodynamic and Maxwell equations that are linearized around
the homogeneous equilibrium state
\begin{eqnarray}
\label{linear} &\pt_\m \d T^{\m\n}=0,&\non &\pt_\m\d
n_{B}^\m=0,&\non &\pt_\m\d n_{e}^\m=0,&\non &\pt_\m \d
H^{\m\n}=0,&\non &\e^{\m\n\a\b}\pt_\n \d F_{\a\b}=0,&\non
\end{eqnarray}
where
\begin{widetext}
\begin{eqnarray}
\label{dtmn} \d T^{\m\n}&=&\d T^{\m\n}_{\rm F0}+\d T_{\rm
EM}^{\m\n}+\d\t^{\m\n},\non\d T^{\m\n}_{\rm F0}&=&\d\ve u^\m
u^\n-\d\pperp\Xi^{\m\n}+\d\ppara b^\m b^\n+(\ve+\pperp)(\d u^\m
u^\n+u^\m\d u^\n)+(\ppara-\pperp)(\d b^\m b^\n+b^\m \d b^\n),\non \d
T_{\rm EM}^{\m\n}&=&B\d B(u^\m u^\n-\Xi^{\m\n}-b^\m b^\n)+B^2(\d
u^\m u^\n+u^\m \d u^\n-\d b^\m b^\n-b^\m\d b^\n),\non \d
\t^{\m\n}&=&\zperp\Xi^{\m\n}\d\f+\zpara b^\m
b^\n\d\vf+\zperp\d\Xi^{\m\n}\f+\zpara\d(b^\m b^\n)\vf,\non \d
n_B^\m&=&\d n_B u^\m+n_B\d u^\m,\non \d n_e^\m&=&\d n_e u^\m+n_e\d
u^\m,\non \d H^{\m\n}&=&\d H b^{\m\n}+H\d b^{\m\n},\non \d
F^{\m\n}&=&\d B b^{\m\n}+B\d b^{\m\n},\non\d\Xi_{\m\n}&=&\d b^\m
b^\n+b^\m\d b^\n-\d u^\m u^\n-u^\m \d u^\n,\non
\d\f&=&\frac{1}{2}\Xi^{\m\n}(\pt_\m \d u_\n+\pt_\n \d u_\m),\non
\d\vf&=&\frac{1}{2} b^\m b^\n(\pt_\m \d u_\n+\pt_\n \d u_\m).
\end{eqnarray}
Upon linearizing the normalization conditions $u^\m u_\m=1,b^\m
b_\m=-1,u^\m b_\m=0$, one finds that the perturbed variables need to
satisfy the constraints
\begin{eqnarray}
\label{constraint} &\d u^\m u_\m=\d b^\m b_\m=u^\m \d b_\m+\d u^\m
b_\m=0.
\end{eqnarray}
In the equations above the perturbations are assumed to have the
form $\d Q=\d Q_0\exp(i k x)$, where $\d Q_0$ is constant, and the
unperturbed quantities are independent of space and time. The
hydrodynamic and Maxwell equations need to be supplemented by an
equation of state (EOS) in order to close the system. The linearized
EOS is given by
\begin{eqnarray}
\label{eos} \d P=c_s^2\d\ve+M\d B,
\end{eqnarray}
where $c_s^2\equiv(\pt P/\pt\ve)_B$ is the speed of sound.

In the most general case \eqs{linear}{eos} constitute 15 independent
equations, but the equations associated with the conservation of
$n_B$ and $n_e$ are decoupled from the others if the EOS is taken in
the form (\ref{eos}). Therefore, we are left with 13 equations.
We work in the rest frame of the equilibrium fluid, $u^\m=(1,0,0,1)$
and $b^\m=(0,0,0,1)$ and choose as independent variables
\begin{eqnarray}
\d Y_i&=&\{ \d u_1,\d u_2,\d u_3,\d b_0,\d b_1,\d b_2,\d \ve,\d
\ppara,\d\pperp,\d B,\d M,\d\f,\d\vf\}.
\end{eqnarray}
(N.B. One can choose other independent variables, but the results
do not change).  The thirteen linear equations can be collected into the
following matrix form
\begin{eqnarray}
G_{ij}\d Y_j=0.
\end{eqnarray}
The matrix $G$ has the following form
\begin{eqnarray}
\noindent &&G=\non&&\non&&\left(
\begin{array}{cccccccccccccc}
 -k_2 & k_1 & 0 & 0 & 0 & 0 & 0 & 0 & 0 & 0 & 0 & 0 & 0 \\
 0 & 0 & 0 & 0 & k_2 & -k_1 & 0 & 0 & 0 & 0 & 0 & 0 & 0 \\
 -Hk_0 & -Hk_0 & 0 & 0 & Hk_3 & Hk_3 & 0 & 0 & 0 & k_1+k_2 & -k_1-k_2 & 0 & 0 \\
 k_3 & k_3 & 0 & 0 & -k_0 & -k_0 & 0 & 0 & 0 & 0 & 0 & 0 & 0 \\
 Bk_1 & Bk_2 & 0 & 0 & 0 & 0 & 0 & 0 & 0 & -k_0 & 0 & 0 & 0 \\
 -hk_1 & -hk_2 & -hk_3 & -HBk_3 & 0 & 0 & k_0 & 0 & 0 & B k_0 & 0 & 0 & 0 \\
 -hk_0 & -hk_0 & 0 & 0 & HBk_3 & HBk_3 & 0 & 0 & k_1+k_2 & B (k_1+k_2) & 0 & -\zperp(k_1+k_2) & 0 \\
 0 & 0 & -hk_0 & -HBk_0 & HBk_1 & HBk_2 & 0 & k_3 & 0 & -B k_3 & 0 & 0 & \zpara k_3 \\
 0 & 0 & 1 & 1 & 0 & 0 & 0 & 0 & 0 & 0 & 0 & 0 & 0 \\
 0 & 0 & 0 & 0 & 0 & 0 & 0 & -1 & 1 & M & B & 0 & 0 \\
 i k_1 & i k_2 & 0 & 0 & 0 & 0 & 0 & 0 & 0 & 0 & 0 & 1 & 0 \\
 0 & 0 & -i k_3 & 0 & 0 & 0 & 0 & 0 & 0 & 0 & 0 & 0 & 1 \\
 0 & 0 & 0 & 0 & 0 & 0 & -c_s^2 & 1 & 0 & -M & 0 & 0 & 0
\end{array}
\right)\non
\end{eqnarray}
where $h=\ve+P+HB$ is the total enthalpy. The exponential plane-wave solutions
for frequencies $k_0$ and wave-vectors $\bk$ satisfy the dispersion relations given by
\begin{eqnarray}
\label{detg} \det G=0.
\end{eqnarray}
For the modes propagating parallel (longitudinal modes) or
perpendicular (transverse modes) to the magnetic field, \eq{detg} has simple
solutions.

1) Transverse modes, $k_3=0$. There are two types of transverse
modes. One is solely determined by the Maxwell equations and has the
following dispersion relation
\begin{eqnarray}
k_0=\pm k_\perp,
\end{eqnarray}
where $k_\perp^2=k_1^2+k_2^2$, and describes simply an electromagnetic wave.
Another solution has the dispersion relation
\begin{eqnarray}
k_0&=&\frac{i \zperp
k_\perp^2\pm\sqrt{4(\ve+\ppara)(\ve+\pperp)c_s^2k_\perp^2-k_\perp^4\zperp^2}}{2(\ve+\ppara)}\non&\approx&\pm\sqrt{\frac{\ve+\pperp}{\ve+\ppara}}c_sk_\perp
+\frac{i \zperp k_\perp^2}{2(\ve+\ppara)},
\end{eqnarray}
where the second approximate relation is valid in the
long-wavelength limit. This solution represents a sound wave
propagating perpendicular to the magnetic field. The speed of this
sonic wave is $\sqrt{(\ve+\pperp)/(\ve+\ppara)} c_s$ and is
smaller than the speed $c_s$ of a ordinary sound wave. It is seen that
positive $\zperp$ implies dissipation of the sonic wave, \ie,  a decay
of the initial disturbance. We conclude that the fluid flow is stable
in the case. However, we see that for negative $\zperp$, the initial
disturbance grows and the fluid is unstable. Thus, we conclude that
negative transverse bulk viscosity implies hydrodynamic instability
via growth of transverse sound waves.

2) Longitudinal modes, $k_1=k_2=0$. We find three types of
longitudinal modes. The first one is again the electromagnetic wave
with the dispersion relation
\begin{eqnarray}
k_0=\pm k_3.
\end{eqnarray}
The second one is a transverse wave oscillating perpendicularly to
the magnetic field, but traveling along the magnetic field lines. It
has the dispersion relation
\begin{eqnarray}
k_0=\pm \sqrt{\frac{BH}{\ve+\ppara+BH}}k_3.
\end{eqnarray}
This mode is the Alfven wave whose speed is equal $\sqrt{BH/(\ve+\ppara+BH)}$.
The third longitudinal mode has the following dispersion relation
\begin{eqnarray}
k_0&=&\frac{i \zpara
k_3^2\pm\sqrt{4(\ve+\ppara)^2c_s^2k_3^2-k_3^4\zpara^2}}{2(\ve+\ppara)}\non&\approx&\pm
c_sk_3 +\frac{i \zpara k_3^2}{2(\ve+\ppara)}.
\end{eqnarray}
This mode represents an ordinary sound wave with dissipation due to the
longitudinal bulk viscosity $\zpara$. It is obvious that if
$\zpara<0$ this mode will not decay, rather grow, thus leading to
hydrodynamic instability.
\end{widetext}
In the next section, we will show that for certain values of the
parameters, the transverse bulk viscosity $\zperp$ could be indeed
negative. We emphasize here that this does not imply
a violation of the second law of thermodynamics, rather this manifests a
hydrodynamic instability of the ground state, \ie, small
perturbations will take the system via this hydrodynamic instability
to a new state.
A candidate state is the one which has inhomogeneous (domain)
structure. Both the structure of the new state and the transition from the
homogeneous to the inhomogeneous state are interesting problems which
are beyond the scope of this study.  However, we would like to point
out a number analogous cases where a negative transport
coefficient indicates instability towards formation of a new state with
domain structure. One such case is the negative resistivity (also known as
the Gunn effect) in certain semiconducting
materials~\cite{gunn:1963,ridley:1963}.
Another case is  the negative (effective) shear viscosity, which is extensively
studied in the literature~\cite{starr,yakhot,cebers,bacri}. Finally,
negative bulk viscosity has been investigated in different contexts
in ~\cite{das:1993,tank:1994}.

%%%%%%%%%%%%%%%%%%%%%%%%%%%%%%%%%%%%%%%%%%%%%%%%%%%%%%%%%%%%%%%%%%%%%%%
\section {Results for the Bulk Viscosities}\label{5}
%%%%%%%%%%%%%%%%%%%%%%%%%%%%%%%%%%%%%%%%%%%%%%%%%%%%%%%%%%%%%%%%%%%%%%%
In order to calculate the bulk viscosities, we need to determine the
coefficients $A, \cpara, \cperp$ and $\l$. From Eqs.
(\ref{pstrong}), (\ref{M0}), and (\ref{nf0}) we obtain in a
straightforward manner
\begin{eqnarray}
\label{coff} A&=&\lb\frac{N_c q_sB}{2\p^2}\sum_{n=0}^{n_{\rm
max}^s}\n_n\frac{\m_s}{\sqrt{\m_s^2-m_s^2-2nq_sB}}\rb^{-1}\non&&+\lb\frac{N_c
q_dB}{2\p^2}\sum_{n=0}^{n_{\rm
max}^d}\n_n\frac{\m_d}{\sqrt{\m_d^2-m_d^2-2nq_dB}}\rb^{-1},\non
\cpara&=&n_{d0}\lb\frac{N_c q_dB}{2\p^2}\sum_{n=0}^{n_{\rm
max}^d}\n_n\frac{\m_d}{\sqrt{\m_d^2-m_d^2-2nq_dB}}\rb^{-1}\non&&-n_{s0}\lb\frac{N_c
q_sB}{2\p^2}\sum_{n=0}^{n_{\rm
max}^s}\n_n\frac{\m_s}{\sqrt{\m_s^2-m_s^2-2nq_sB}}\rb^{-1},\non
\cperp&=&\cpara-\lb\frac{\pt M}{\pt\m_d}/\frac{\pt
n_{d0}}{\pt\m_d}-\frac{\pt M}{\pt\m_s}/\frac{\pt n_{s0}}{\pt\m_s}\rb
B,
\end{eqnarray}
where
\begin{eqnarray}
\frac{\pt M}{\pt\m_f}&=&\frac{N_c q_f}{2\p^2}\sum_{n=0}^{n_{\rm
max}^f}\n_n\frac{\m_f^2-m_f^2-3nq_fB}{\sqrt{\m_f^2-m_f^2-2nq_fB}}.
\end{eqnarray}

The rate $\l$ of the weak processes (\ref{weak1}) and (\ref{weak2})
should also be affected by a strong magnetic field. The major effect
of a magnetic field on $\l$ is to modify the phase space of weak
processes (\ref{weak1}) and
(\ref{weak2})~\cite{Chakrabarty:1998su,XiaoPing:2005gw}. Taking this
into account, one obtains
\begin{eqnarray}
\label{lambda} \l&=&\frac{64\p^5}{5} \tilde{G}^2\m_d
T^2\non&&\times\lb \frac{q_u B}{2\p^2}\sum_{n=0}^{n^u_{\rm
max}}\n_n\frac{1}{\sqrt{\m_u^2-m_u^2-2n q_u B}}\rb^2\non&&\times\lb
\frac{q_d B}{2\p^2}\sum_{n=0}^{n^d_{\rm
max}}\n_n\frac{1}{\sqrt{\m_d^2-m_d^2-2n q_d B}}\rb\non&&\times\lb
\frac{q_s B}{2\p^2}\sum_{n=0}^{n^s_{\rm
max}}\n_n\frac{1}{\sqrt{\m_s^2-m_s^2-2n q_s B}}\rb.
\end{eqnarray}
where $\tilde{G}^2\equiv
G_F^2\sin^2\h_C\cos^2\h_C=6.46\times10^{-24}{\rm MeV}^{-4}$ is the
Fermi constant.

\begin{figure}[!htb]
\begin{center}
\includegraphics[width=7.5cm]{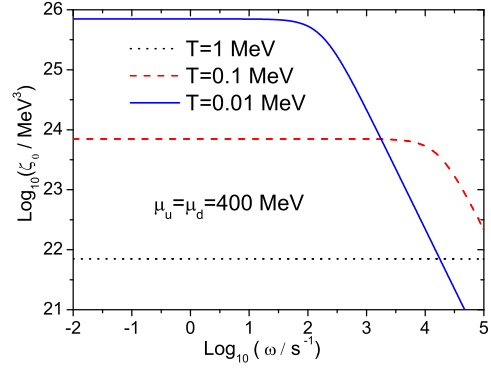}
\caption{(Color online) The isotropic bulk viscosity $\z_0$ at zero
magnetic field as function of the oscillation frequency $\o$ for
$\mu_u = \mu_d = 400 $ MeV at $T=0.01$ (solid, blue online) $0.1$
(dashed, red online) and $1$ (black, dotted online) MeV. }
\label{zeta0}
\end{center}
\end{figure}

1) When the magnetic field is much smaller than the typical chemical
potential, say, $q_d B\ll \m_d^2$, its effect on the bulk
viscosities is negligible. For typical parameters (\ref{paras}),
this condition holds up to $B\sim 10^{17}$ G. In this case, the
system is practically isotropic, $\zpara$ and $\zperp$ are
effectively degenerate with the isotropic $\z_0$, the bulk viscosity
of unmagnetized matter. The zero-magnetic field limit results can be
obtained easily by replacing  $\sum_{n=0}^{n_{\rm max}^f} q_f
B\ra2k_{Ff}^2,$ and $B\ra0$. The bulk viscosity for zero magnetic
field $\z_0$ as function of oscillation frequency $\o$ for various
temperature is shown in \fig{zeta0}. The ``shoulder" structure and
the temperature dependence of $\z_0$ are easily understood from
\eq{vis1} and have been widely discussed in the
literature~\cite{Madsen:1992sx,Zheng:2002jq,Xiaoping:2005js,Anand:1999bj,Sad:2006qv,Alford:2006gy,Sad:2007ud,Dong:2007mb,Alford:2007rw,Manuel:2007pz,Dong:2007ax,Alford:2008pb,arXiv:0806.1005}.
%%%%%%%%%%%%%%%%%%%%%%%%%%%%%%%%%%%%%%%%%%%%%%%%%%%%%%%%%%%%%%%%%%%%%%%
%%%%%%%%%%%%%%%%%%%%%%%%%%%%%%%%%%%%%%%%%%%%%%%%%%%%%%%%%%%%%%%%%%%%%%%

2) When the magnetic field is extremely large, say, $B\gg
B_c\sim10^{20}$ G, for our choice of parameters (\ref{paras}) all
the quarks are confined in their lowest Landau level. In this case
we obtain
\begin{eqnarray}
A&=&\frac{2\p^2}{N_c q_d\m_d B}(k_{Fd}+k_{Fs}),\non
\cpara&=&\frac{m_s^2-m_d^2}{\m_d},\non
\cperp&=&0,\non\l&=&\frac{4G^2q_u^2q_d^2B^4T^2}{5\p^3\m_u^2 k_{Fs}},
\end{eqnarray}
and therefore
\begin{eqnarray}
\zperp&=&0,\non\zpara&\approx&\frac{45 m_s^4\m_u^2 k_{Fs}}{16 \p
\tilde{G}^2 q_u^2 B^2 T^2 (k_{Fs}+k_{Fd})^2}.
\end{eqnarray}
We used the parameters (\ref{paras}) and assumed physically
interesting frequencies $\o<10^4$ s$^{-1}$.  The bulk viscosity
$\zperp$  vanishes as a  consequence of vanishing $\pperp$ when
$B>B_c$. Since $\zpara$ is now inversely proportional to $B^2$ it
approaches zero for large $B$. Therefore, both $\zperp$ and $\zpara$
are suppressed for large $B$. In contrary, $\ppara$ is enhanced by
the extremely large magnetic field, see \fig{pressure_u}.

\begin{figure}[!htb]
\begin{center}
\includegraphics[width=7.5cm]{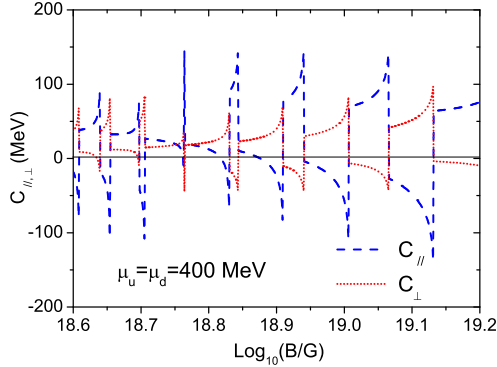}
\caption{(Color online) Coefficients $\cpara$ (dashed, blue online)
and $\cperp$ (dotted, red online) as functions of $B$ at zero
temperature.} \label{c_u}
\end{center}
\end{figure}

3) When the magnetic field is strong, but not strong enough to
confine all the quarks to their lowest Landau level, the situation
becomes complicated. For our chosen parameters (\ref{paras}), this
situation  roughly corresponding to the interval $10^{17} {\rm
G}<B<10^{20}$ G. In this case, a finite number of Landau levels is
occupied, and the essential observation is that $\cpara$ and
$\cperp$ can be negative. The behaviors of $\cpara$ and $\cperp$ are
shown in \fig{c_u} as functions of $B$. Let us concentrate on the
few levels just above the value $10^{19}$ G. When $B$ grows passing
over $B_n^d$ or $B_n^s$ for each $n$, both $\cpara$ and $\cperp$
change their sign. More importantly, they have always opposite
signs. Therefore, in this region, $\zperp$ is negative which leads
to hydrodynamic instability (see the analysis in Sec.\ref{43}).
%%%%%%%%%%%%%%%%%%%%%%%%%%%%%%%%%%%%%%%%%%%%%%%%%%%%%%%%%%%%%%%%%%%%%%%
%%%%%%%%%%%%%%%%%%%%%%%%%%%%%%%%%%%%%%%%%%%%%%%%%%%%%%%%%%%%%%%%%%%%%%%

\begin{figure}[!htb]
\begin{center}
\includegraphics[width=9cm]{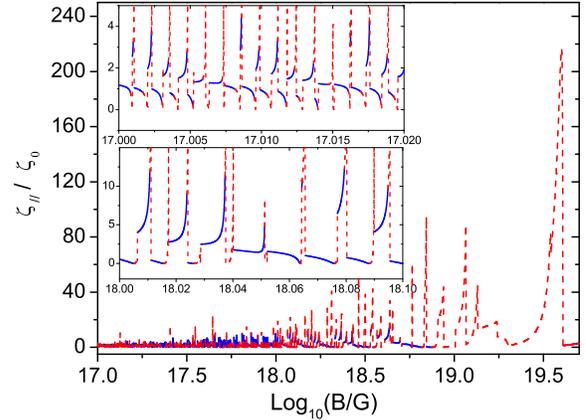}
\includegraphics[width=9cm]{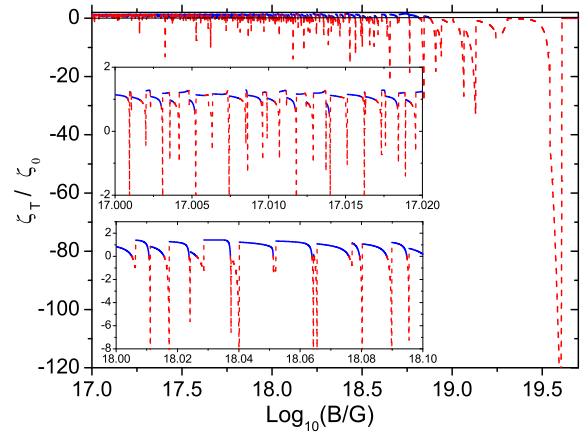}
\caption{(Color online) Bulk viscosities $\zpara$ and $\zperp$
scaled by the isotropic bulk viscosity $\z_0$ as functions of the
magnetic field $B$ at fixed frequency $\o=2\p\times \,10^3\, {\rm
s}^{-1}$ and temperature $T=0.1$ MeV. The dashed red curves
correspond to viscosities lying in the unstable regions and would be
not physically reachable. Sub-panels show the amplifications around
$10^{17}$ and $10^{18}$ G. Our parameters are given in \eq{paras}.}
\label{zeta_u}
\end{center}
\end{figure}

The numerical values of the bulk viscosities $\zpara$ and $\zperp$
are shown in \fig{zeta_u} as functions of $B$. The parameters are
those given in \eq{paras}. We also fix the temperature $T=0.1$ MeV
and oscillation frequency $\o=2\p\times 10^3 {\rm s}^{-1}$. Both
$\zpara$ and $\zperp$ have ``quasi-periodic" oscillatory dependence
on the magnetic field. The two boundaries of each ``period"
correspond to a pair of neighboring $B_n^f,\,f=u,d,s$ and
$n=0,1,2\cdots$, and hence the period is roughly $\D
B\sim2q_fB^2/k^2_{Ff}$ for large $B$. Therefore, on average, the
period increases as $B$ grows. The amplitude of these oscillations
also grows with increasing magnetic field until $B\simeq B_c$.
Thereafter all the quarks are confined to their lowest Landau levels
and $\zperp$ vanishes. From \fig{zeta_u} we see that the magnitudes
of $\zperp$ and $\zpara$ can be 100 to 200 times larger than their
zero field value $\z_0$. Due to the unequal masses and charges of
$u,d$ and $s$ quarks, $\zpara$ and $\zperp$ behave very irregularly.
We illustrate the zoomed-in curves around $B=10^{17}$ and
$B=10^{18}$ G in the sub-panels, which look more regular. The
quasi-periodic structures are more evident in these sub-panels.
%%%%%%%%%%%%%%%%%%%%%%%%%%%%%%%%%%%%%%%%%%%%%%%%%%%%%%%%%%%%%%%%%%%%%%%
%%%%%%%%%%%%%%%%%%%%%%%%%%%%%%%%%%%%%%%%%%%%%%%%%%%%%%%%%%%%%%%%%%%%%%%

The most unusual feature seen in \fig{zeta_u} is that for a wide
range of field values, the transverse bulk viscosity $\zperp$ is
negative. Therefore, strange quark matter in this region is
hydrodynamically unstable. Besides this hydrodynamical instability,
near each $B_n^f$, there is a narrow window in which thermodynamical
instability arises. We depict $\zpara$ and $\zperp$ in these
unstable regions by dashed red curves. The solid blue curves
correspond to the stable regime.

\begin{figure}[!htb]
\begin{center}
\includegraphics[width=7.5cm]{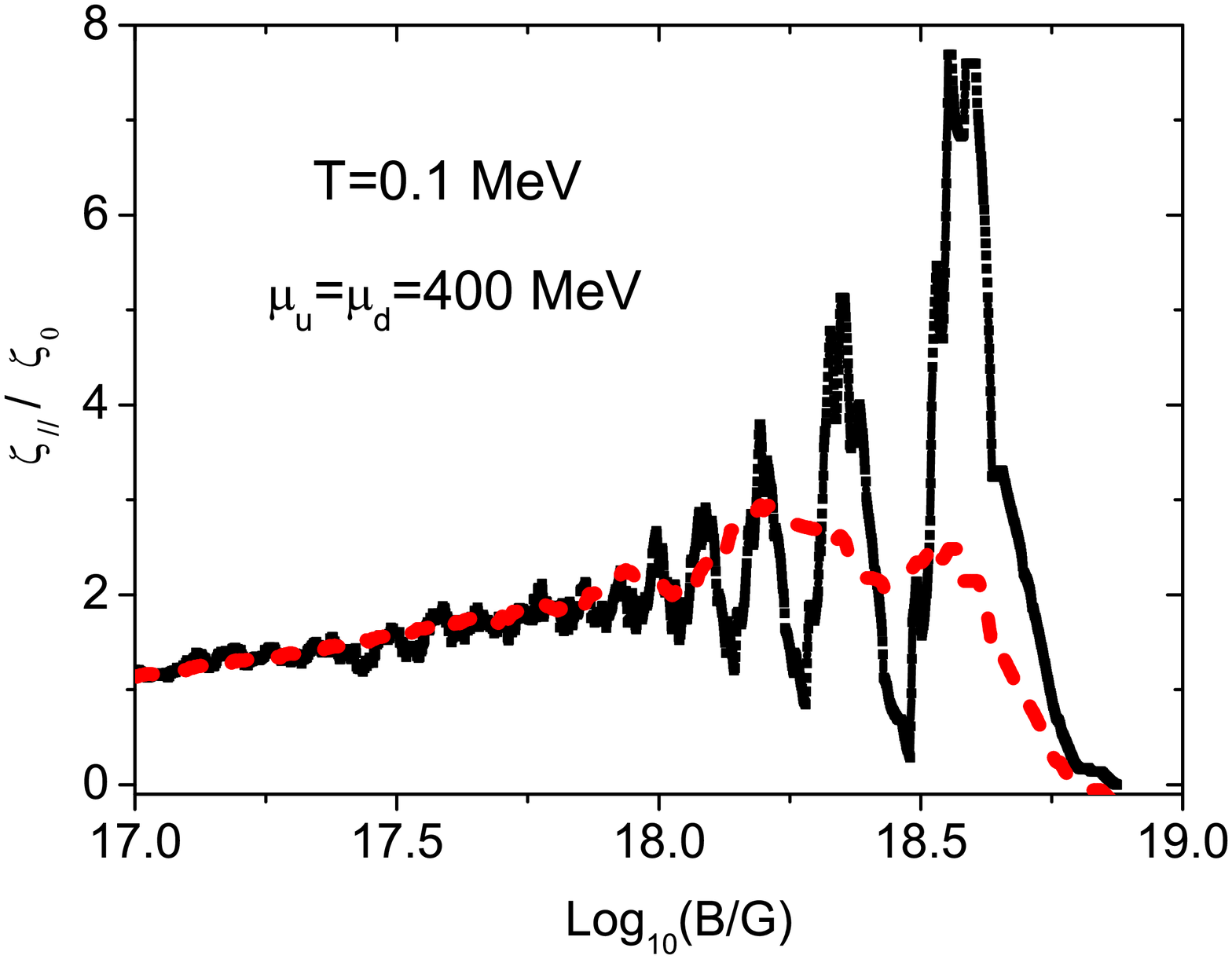}
\includegraphics[width=7.5cm]{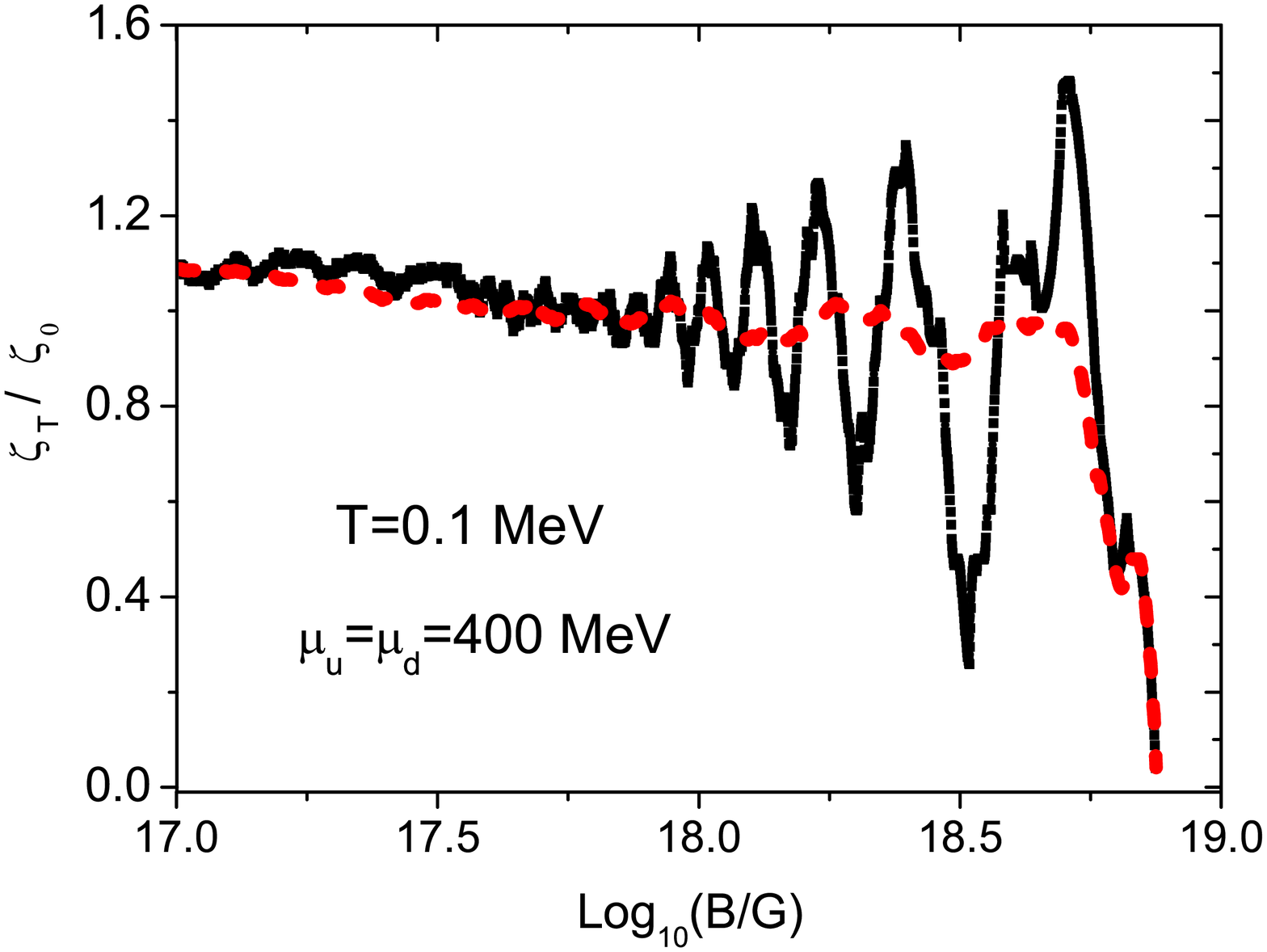}
\caption{(Color online) The averaged bulk viscosities $\zpara$ and
$\zperp$ scaled by the isotropic bulk viscosity $\z_0$ as functions
of the magnetic field $B$ at fixed frequency $\o=2\p\times \,10^3\,
{\rm s}^{-1}$ and temperature $T=0.1$ MeV. The black-solid curve
corresponds to averaging over a short period
($\D\log_{10}(B/G)=0.05$); while the red short-dashed curve
corresponds to averaging over a long period
($\D\log_{10}(B/G)=0.5$). The viscosities lying in the unstable
regions have been eliminated in the averaging.} \label{zetaaver}
\end{center}
\end{figure}

The magnetic field in a compact star need not be homogeneous and may
have a complicated structure with poloidal and toroidal components.
Furthermore, the fields will be functions of position in the star
because of the density dependence of the parameters of the theory.
Furthermore, the instabilities, described above, may lead to
fragmentation of matter and formation of domain structures, where
the regions with magnetic fields are separated from those without
magnetic field by domain walls. Accordingly, only the averaged
viscosities over some range of magnetic field have practical sense
for assessing the large-scale behavior of matter. Averaging over
many oscillation periods in the stable region, we find that the
averaged values of $\zpara$ and $\zperp$ are much more regular, with
their magnitudes restricted from 0 to several $\z_0$, see
\fig{zetaaver}. In obtaining the curves in \fig{zetaaver}, we have
eliminated the viscosities lying in the unstable regime. The solid
black  curves are obtained by averaging over a short period
$\D\log_{10}(B/G)=0.05$. The period was chosen such that the most
rapid fluctuations are smeared out, but the oscillating structures
over larger scale are intact. The short-dashed red curves correspond
to averaging over an even longer period, $\D\log_{10}(B/G)=0.5$. The
result of long-period averaging is that $\zpara$ first increases
slowly and then drops down quickly once $B>10^{18.5} $G; similarly,
$\zperp$ first slowly decreases and then drops down very fast for
$B\sim 10^{18.5} $ G. Such a dropping behavior reflects the fact
that a large number of quarks are beginning to occupy the lowest
Landau level.
%%%%%%%%%%%%%%%%%%%%%%%%%%%%%%%%%%%%%%%%%%%%%%%%%%%%%%%%%%%%%%%%%%%%%%%
%%%%%%%%%%%%%%%%%%%%%%%%%%%%%%%%%%%%%%%%%%%%%%%%%%%%%%%%%%%%%%%%%%%%%%%

We note that the appearances of thermodynamic, mechanical, and
hydrodynamical instabilities are all induced by the Landau
quantization of the quark levels, \ie, are quantum mechanical in
nature. More precisely, they are all due to the interplay between
the Landau levels and the Fermi momentum (reflected in the quantity
$B_n^f$). Additionally, the hydrodynamical instability requires that
the quark matter is para-magnetized. Although we did our analysis by
using the free quark gas approximation, \eq{pstrong}, it should be
valid
 as long as there are sharp
Fermi surfaces (low temperature), quantized Landau levels
(high magnetic field) and para-magnetization. The
appearances of these instabilities are expected to be robust
feature for such systems.

We also checked that if one imposes the neutrality condition, \ie,
the condition $2n_u=n_d+n_s$, there is only minor quantitative
change, while the qualitative conclusions are almost unchanged. Our
choice of chemical potential $\m_u=\m_d=400$ MeV  roughly
corresponds to the choice of $n_B\sim 4-5 n_0$ for neutral strange
quark matter.

%%%%%%%%%%%%%%%%%%%%%%%%%%%%%%%%%%%%%%%%%%%%%%%%%%%%%%%%%%%%%%%%%%%%%%%
\section {R-Mode Instability Window}\label{6}
%%%%%%%%%%%%%%%%%%%%%%%%%%%%%%%%%%%%%%%%%%%%%%%%%%%%%%%%%%%%%%%%%%%%%%%

The purpose of this section is to discuss the damping of the r-modes
of Newtonian models of strange stars by dissipation driven by the
bulk viscosities $\zperp$ and $\zpara$. As is well known, rotating
equilibrium configurations of self-gravitating fluids are
susceptible to instabilities at high rotation rates. Starting from
the mass-shedding limit and going down with the rotation rate, the
first instability point corresponds to the dynamical instability of
the $l=2$ and $m=2$ mode. This bar-mode instability is independent
of the dissipative processes inside the star and occurs at values of
the kinetic to potential energy ratio  $T/W\sim
0.27$~\cite{Saijo:2000qt}. For smaller rotation rates two secular
instabilities with $l=2$ arise, each corresponding to a sign of
$m=\pm 2$. For incompressible fluids at constant density the $T/W$
values for the onset of secular instabilities coincide. One
instability is driven by the viscosity, the other instability is
driven by the gravitational radiation. For realistic stars the $T/W$
values for the onset of these instabilities do not coincide;
relativity and other factors shift the viscosity-driven instability
to higher values of $T/W$. At the same time the gravitational
radiation instability is shifted to lower values of $T/W$. The
gravitational radiation instability arises for the modes which are
retrograde in the co-rotating frame, while prograde in the (distant)
laboratory frame. The underlying mechanism is the well established
Chandrasekhar-Friedman-Schutz (CFS)
mechanism~\cite{Chandrasekhar:1970,Friedman:1978hf}. The bulk and
shear viscosities can prevent the development of the CFS
instabilities, except in a certain window in the rotation and
temperature plane.

In the following we shall concentrate on axial modes of Newtonian
stars, the so-called r-modes, which are known to undergo a CFS-type
instability. Our main goal will be to assess the role of strong
magnetic fields and bulk viscosity on the stability of these
objects. We shall adopt the formalism of
Refs.~\cite{Ipser:1990,Lindblom:1999yk,gr-qc/0010102} for our study
of the damping of the r-modes by bulk viscosity. For the sake of
simplicity we shall describe both fluid mechanics and gravity in the
Newtonian approximation.

The equations that describe the dynamical evolution of the star are
\begin{subequations}
\begin{eqnarray}
\label{newtonian1} &\pt_t\r+\nabla\cdot(\r\bv)=0,&
\end{eqnarray}
\begin{eqnarray}
\label{newtonian2}
&\pt_t\bv+\bv\cdot\nabla\bv=-\nabla(h-\F)\equiv-\nabla U,&
\end{eqnarray}
\begin{eqnarray}
\label{newtonian3} &\nabla^2\F=-4\p G \rho,&
\end{eqnarray}
\end{subequations}
where $h$ is defined by the integral
\begin{eqnarray}
h(P)\equiv\int^P_0\frac{dP'}{\r(P')}.
\end{eqnarray}
The quantity $\r$ is the mass density of the fluid which is assumed
to satisfy a barotropic equation of state, $\r=\r(P)$. $\F$ is the
gravitational potential and $G$ is the gravitational constant. The
potential $U$ is used to determine the velocity field $\bv$.

The oscillation modes of a uniformly rotating star can be completely
described in terms of two perturbation potentials $\d U\equiv U-U_0$
and $\d\F\equiv\F-\F_0$, where $U_0$ and $\F_0$ are the potentials
that correspond to the equilibrium configuration of the star. We
assume that the time and azimuthal angular dependence of any
perturbed quantity is described by $\propto e^{i\ot t+im\vf}$, where
$m$ is an integer and $\ot$ is the frequency of the mode in
laboratory frame. Let $\O$ denote the rotation frequency of the star
and $\o$ denote the frequency of the perturbed quantity measured in
the co-rotating frame [which corresponds the $\o$ in Eqs.
(\ref{vis1}) and (\ref{vis2}) because we will work in the
co-rotating frame]. For small $\O$ there is a simple relation
between $\O, \ot$, and $\o$~\cite{Ipser:1990,Lindblom:1999yk},
\begin{eqnarray}
 \o=\ot+m\O.
\end{eqnarray}
By linearizing the Euler equation
(\ref{newtonian2}) around the equilibrium configuration, the velocity
perturbation $\d v^a$ is determined by~\cite{Ipser:1990,Lindblom:1999yk}
\begin{eqnarray}
\label{dv} \d v^a=i Q^{ab}\nabla_b\d U.
\end{eqnarray}
The tensor $Q^{ab}$ is a function of $\ot$ and the rotation
frequency $\O$ of the star,
\begin{eqnarray}
Q^{ab}&&\!\!\!\!\!=\frac{1}{(\ot+m\O)^2-4\O^2}\non&&\!\!\!\!\!\times\ls(\ot+m\O)\d^{ab}-\frac{4\O^2}{\ot+m\O}z^a
z^b-2i\nabla^a v^{b}_0\rs,\;\;\;
\end{eqnarray}
where $\bz$ is a unit vector pointing along the rotation axis of the
equilibrium star, which we assume to be parallel to the magnetic field,
\ie, $z_i=b_i$ in Cartesian coordinate system. Here
$\bv_0=r\O \sin\h \hat{\vf}$ is the fluid velocity of the equilibrium
star.

Having the linearized Euler equation, one proceeds to the
linearization of the mass continuity equation (\ref{newtonian1}) and
the equation for the gravitational potential (\ref{newtonian3}); one
finds
\begin{eqnarray}
\label{newtonian4} &\nabla_a(\r Q^{ab}\nabla_b\d U)=-(\ot+m\O)(\d
U+\d \F)d\r/{d h},&\non &\nabla^2\d\F=-4\p G (\d U+\d \F)d\r/{d h}.&
\end{eqnarray}
These equations, together with the appropriate boundary conditions
at the surface of the star for $\d U$ and at infinity for $\d \F$,
determine the potentials $\d U$ and $\d \F$.

For slowly rotating stars, \eq{newtonian4} can be solved order by
order in $\O$,
\begin{eqnarray}
\label{dudf} \d U=R^2\O^2\ls\d U_0+\d U_2\frac{\O^2}{\p G
\r}+O(\O^4)\rs, \non\d \F=R^2\O^2\ls\d \F_0+\d \F_2\frac{\O^2}{\p G
\r}+O(\O^4)\rs,
\end{eqnarray}
where $R$ is the radius of nonrotating star. Since we need only the
perturbed velocity, we will focus on $\d U$ in the following
discussion. The zeroth-order contribution to the r-mode is generated
by the following form of the potential $\d U_0$,
\begin{eqnarray}
\d U_0&=&\a\lb\frac{r}{R}\rb^{m+1}P^m_{m+1}(\cos\h)e^{i\ot t+im\vf},
\end{eqnarray}
\begin{eqnarray}
\label{omega} \ot&=&-\frac{(m-1)(m+2)}{m+1}\O,
\end{eqnarray}
where $\a$ is an arbitrary dimensionless constant and $P_l^m(x)$ are
the associated Legendre polynomials. It has been shown that the most
unstable mode is the one with
$m=2$~\cite{gr-qc/0010102,gr-qc/9803053}, therefore we shall
consider only this case in the following discussion. Substituting
$\d U_0$ into \eq{dv} one obtains the first-order perturbed
velocity,
\begin{eqnarray}
\label{dv0} \d {\bf v}_0&=&\a' R\O\lb\frac{r}{R}\rb^{m}{\bf
Y}^B_{mm}(\h,\vf)e^{i\o t},
\end{eqnarray}
where $\a'=\a\sqrt{\p(m+1)^3(2m+1)!/m}$ and ${\bf Y}^B_{lm}(\h,\vf)$
is the magnetic-type spherical harmonic function,
\begin{eqnarray}
{\bf Y}^B_{lm}(\h,\vf)=\frac{\br\times\nabla Y_{lm}}{\sqrt{l(l+1)}}.
\end{eqnarray}

It is straightforward to check that the first-order perturbed
velocity satisfies
\begin{eqnarray}
\frac{\pt \d v_{0z}}{\pt z}&=&0,\non \nabla\cdot\d {\bf v}_{0}&=&0,
\end{eqnarray}
therefore it does not contribute to the dissipation due to the bulk
viscosities $\zperp$ and $\zpara$. In order to see how the bulk
viscosities damp the r-mode instability one must consider
next-to-first order, \ie, the third-order perturbed velocity which
is generated by the potential $\d U_2$. One can not determine
analytically $\d U_2$ from \eqs{newtonian4}{dudf}, but the angular
structure of $\d U_2$ can be well represented by the following
spherical harmonics expansion ~\cite{Lindblom:1999yk},
\begin{eqnarray}
\d U_2&=&\a f_1(r)P_{m+1}^1(\cos\h)e^{i\ot t+im\vf}\non &&+\a
f_2(r)P_{m+3}^m(\cos\h)e^{i\ot t+im\vf}.
\end{eqnarray}
The functions $f_1(r)$ and $f_2(r)$ have been determined
numerically in Ref.~\cite{Lindblom:1999yk}. A useful approximation
is provided by the following simple expressions
\begin{widetext}
\begin{eqnarray}
\label{f1}
f_1(r)&=& -0.1294\lb\frac{r}{R}\rb^3 -
0.0044\lb\frac{r}{R}\rb^4 + 0.1985\lb\frac{r}{R}\rb^5 -
0.0388\lb\frac{r}{R}\rb^6, \\
\label{f2} f_2(r)&=&-0.0092\lb\frac{r}{R}\rb^3 +
0.0136\lb\frac{r}{R}\rb^4 - 0.0273\lb\frac{r}{R}\rb^5 -
0.0024\lb\frac{r}{R}\rb^6,
\end{eqnarray}
\end{widetext}
which excellently fit the numerical result. We will
use Eqs. (\ref{f1}) and (\ref{f2}) in the following numerical
calculation.

The energy of r-modes comes both from the velocity perturbation and
the perturbation of the gravitational potential. For slowly rotating
stars, the main contribution comes from the velocity
perturbation~\cite{gr-qc/0010102,astro-ph/0101136,gr-qc/9803053,gr-qc/9804044}.
Then, the energy of the r-mode measured in the co-rotating frame is
\begin{eqnarray}
\tilde{E}=\frac{1}{2}\int \r \d\bv^*\cdot\d\bv d^3\br.
\end{eqnarray}
Assuming spherical symmetry, we have
\begin{eqnarray}
\tilde{E}=\frac{1}{2}\a'^2\O^2R^{-2m+2}\int_0^R \r r^{2m+2} dr.
\end{eqnarray}
This energy will be dissipated both by gravitational radiation and
by the thermodynamic transport in the
fluid~\cite{gr-qc/0010102,astro-ph/0101136},
\begin{eqnarray}
\frac{d\tilde{E}}{dt}&=&\lb\frac{d\tilde{E}}{dt}\rb_{G}+\lb\frac{d\tilde{E}}{dt}\rb_{T}.
\end{eqnarray}
The dissipation rate due to gravitational radiation
is given by~\cite{gr-qc/0010102,astro-ph/0101136,arXiv:0806.3359}
\begin{eqnarray}
\lb\frac{d\tilde{E}}{dt}\rb_{G}&&\!\!\!\!\!\!\!\!\!\!=-\ot(\ot+m\O)\sum_{l\geq
2} N_l\o^{2l}\ls|\d D_{lm}|^2+|\d J_{lm}|^2\rs,\non
\end{eqnarray}
where
\begin{eqnarray}
N_l=\frac{4\p G(l+1)(l+2)}{l(l-1)[(2l+1)!!]^2}.
\end{eqnarray}
$\d D_{lm}$ and $\d J_{lm}$ are the mass and current multipole
moments of the perturbation,
\begin{eqnarray}
\d D_{lm}&=&\int\d\r r^l Y_{lm}^*d^3\br,\non \d
J_{lm}&=&2\sqrt{\frac{l}{l+1}}\int r^l(\r\d\bv+\d\r\bv)\cdot{\bf
Y}_{lm}^{B*} d^3\br.
\end{eqnarray}
Taking into account \eq{omega} one obtains
\begin{eqnarray}
\ot(\ot+m\O)=-\frac{2(m-1)(m+2)}{(m+1)^2}\O^2<0,
\end{eqnarray}
which implies that the total sign of $(d\tilde{E}/dt)_{G}$ is
positive: gravitational radiation always increases the energy of
the r-modes.

In order to compare the relative strengths of different dissipative
processes, it is convenient to introduce the dissipative timescales
defined by
\begin{eqnarray}
\t_i\equiv- \frac{2\tilde{E}}{(d\tilde{E}/dt)_i},
\end{eqnarray}
where the index $i$ labels the dissipative process.

The lowest-order contribution to $(d\tilde{E}/dt)_G$ comes from the
current multipole moment $\d J_{ll}$. For the most important case
$l=m=2$, this leads to the following timescale (derived for a simple
polytropic equation of state
$P\propto\r^2$)~\cite{gr-qc/9803053,Lindblom:1999yk}
\begin{eqnarray}
{1\over\t_G}=-\frac{1}{3.26}\lb\frac{\O^2}{\p G\r}\rb^3 s^{-1}.
\end{eqnarray}

The bulk viscosities $\zperp$ and $\zpara$ dissipate the energy of the
r-mode according to
\begin{eqnarray}
\lb\frac{d\tilde{E}}{dt}\rb_\zperp&=&-\frac{3}{2}\int\zperp\bigg|\frac{\pt\d
v_x}{\pt x}+\frac{\pt\d v_y}{\pt y}\bigg|^2
d^3\br,\non\lb\frac{d\tilde{E}}{dt}\rb_{\zpara}&=&-3\int\zpara\bigg|\frac{\pt\d
v_z}{\pt z}\bigg|^2 d^3\br.
\end{eqnarray}
Accordingly, the time scales $\t_\zperp$ and $\t_{\zpara}$ are given
by
\begin{eqnarray}
\t_{\zperp,\zpara}&=&-
\frac{2\tilde{E}}{(d\tilde{E}/dt)_{\zperp,\zpara}}.
\end{eqnarray}

Currently, the shear viscosities $\w_1-\w_5$ of strange quark matter
are not known. In order to determine the damping of the r-mode by
shear viscosity, we take as a crude estimate the value of $\w_0$  in
the absence of a magnetic field~\cite{Heiselberg:1993cr}
\begin{eqnarray}\label{eta0}
\w_0\simeq\w&=&5.5\times10^{-3}\a_s^{-5/3}\m_d^{14/3}\times
T^{-5/3},
\end{eqnarray}
where $\a_s$ is the coupling constant of strong interaction. We will
choose the value $\a_s=0.1$ and apply Eq. (\ref{eta0}) to highly degenerate
3-flavor quark matter with equal chemical potentials of all flavors,
 ($\m_u\simeq\m_d\simeq\m_s$). The contribution to the energy
dissipation rate $\tilde{E}$  due to shear
viscosity $\w$ now becomes
\begin{eqnarray}
\lb\frac{d\tilde{E}}{dt}\rb_\w&=&-\int\w|w_{ij}-\d_{ij}\h/3|^2
d^3\br.
\end{eqnarray}
Assuming a uniform mass density star, the time scale $\t_\w$ can be
simply expressed as~\cite{arXiv:0806.3359}
\begin{eqnarray}
\frac{1}{\t_\w}=\frac{7\w}{\r R^2}.
\end{eqnarray}

The total time scale $\t(\O, T)$ is given  by the following
sum
\begin{eqnarray}
\frac{1}{\t}\equiv\frac{1}{\t_G}+\frac{1}{\t_\zperp}+\frac{1}{\t_\zpara}+\frac{1}{\t_\w},
\end{eqnarray}
which characterizes how fast the r-mode decays. Most
importantly, if the sign of $\t$ is negative the amplitude
of the r-mode will not decay, rather it will increase with time.
Thus, it is important to determine the
critical angular velocity $\O_c$ for the onset of instability
\begin{eqnarray}
\frac{1}{\t(\O_c,T)}=0.
\end{eqnarray}
At a given temperature, stars with $\O>\O_c$ will be unstable due to
gravitational radiation.

%%%%%%%%%%%%%%%%%%%%%%%%%%%%%%%%%%%%%%%%%%%%%%%%%%%%%%%%%%%%%%%%%%%%%%%
\begin{figure}[!htb]
\begin{center}
\includegraphics[width=7.5cm]{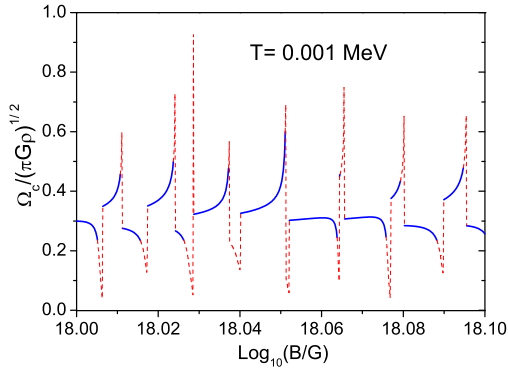}
\caption{(Color online) The critical angular velocity $\O_c$ of a
strange quark star as a function of magnetic field $B$ at
temperature $T=0.001$ MeV. The red dashed lines correspond to the
unstable, while the blue solid lines correspond to the stable
regime.} \label{rB}
\end{center}
\end{figure}
\begin{figure}[!htb]
\begin{center}
\includegraphics[width=6.5cm]{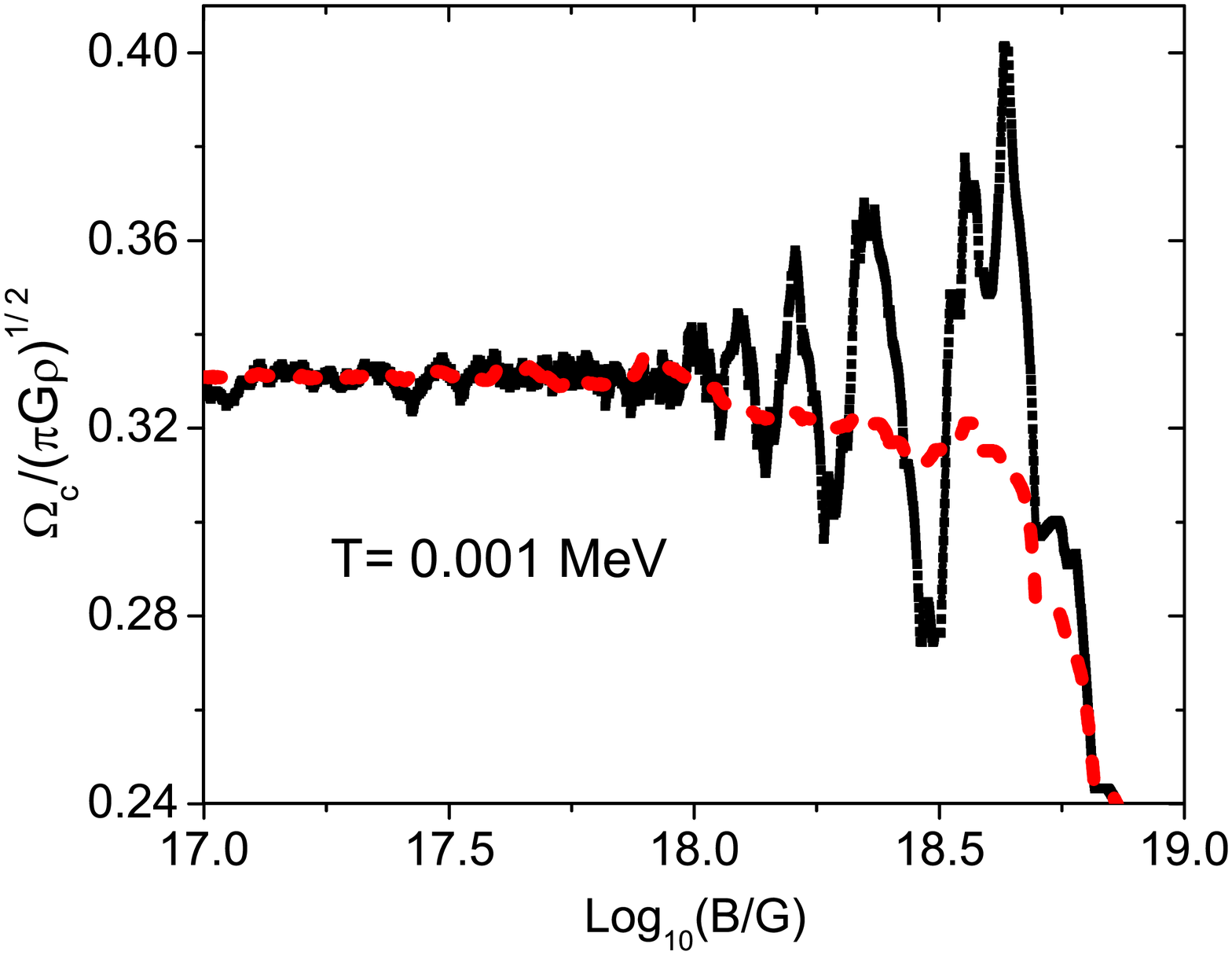}
\includegraphics[width=6.5cm]{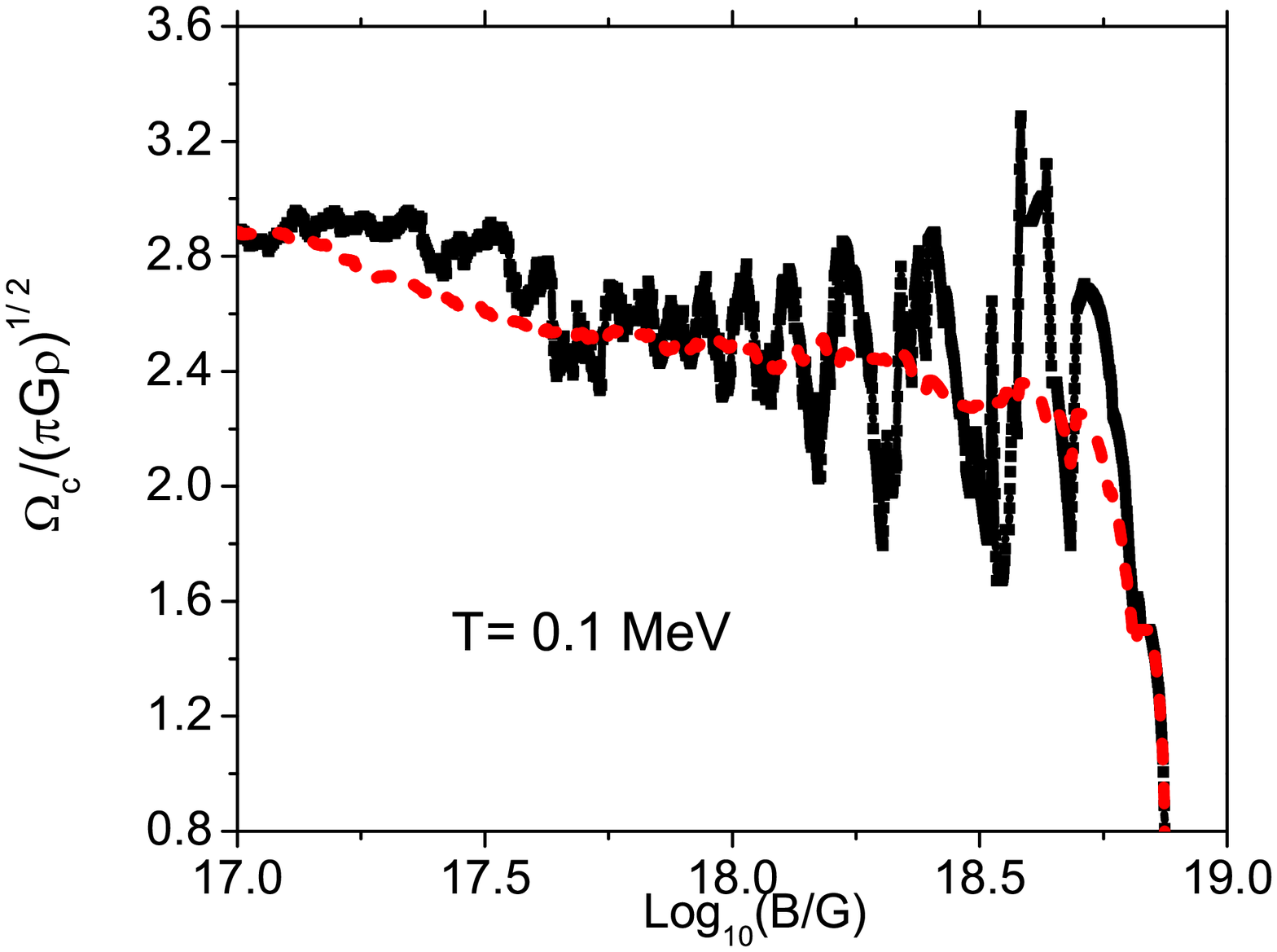}
\includegraphics[width=6.5cm]{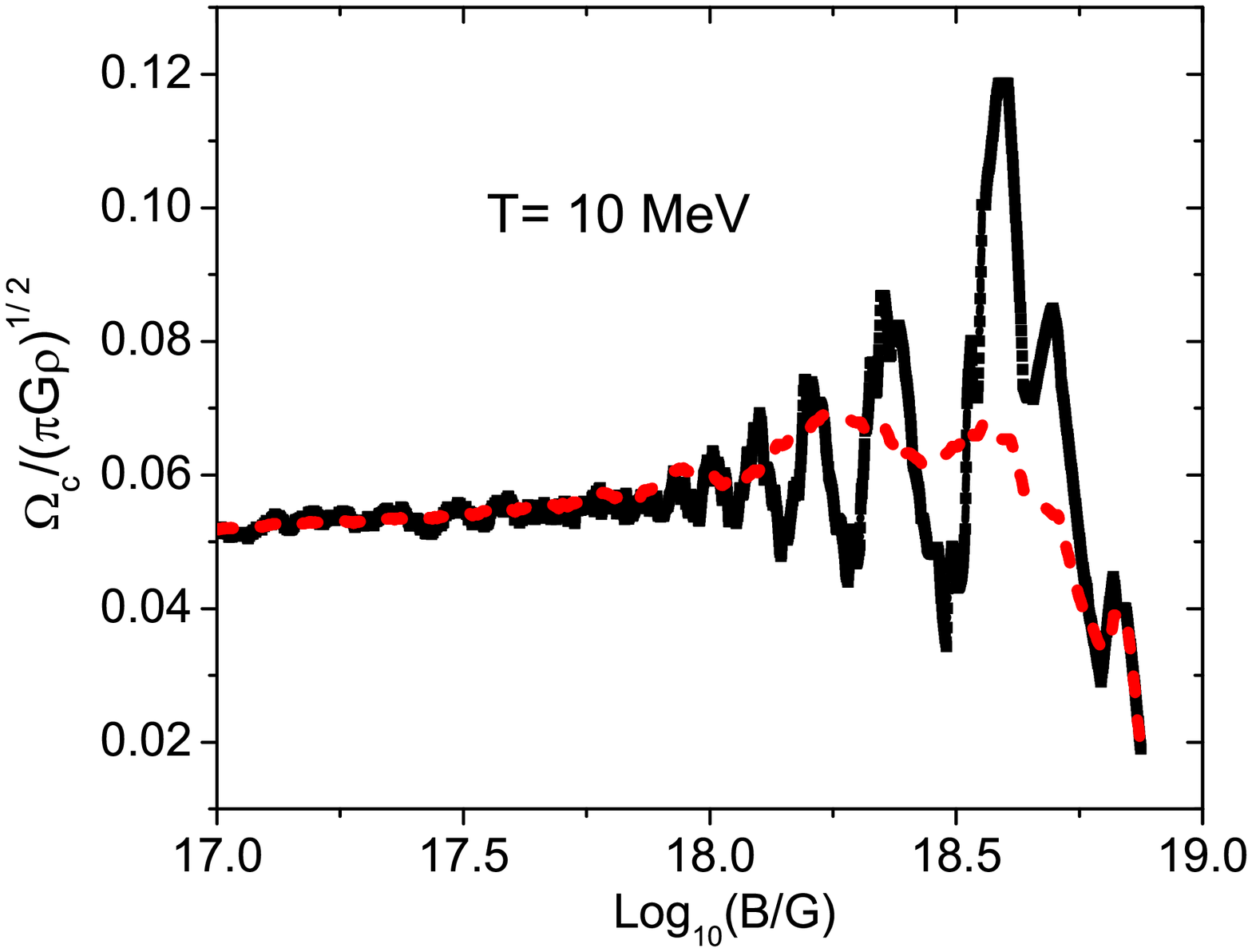}
\caption{(Color online) The averaged critical angular velocity
$\O_c$ of a strange quark star as a function of magnetic field $B$
at various temperatures $T=0.001,\; 0.1$, and $10$ MeV. The black
solid curve corresponds to averaging over a short period
($\D\log(B/G)=0.05$); while the red short-dashed curve corresponds
to averaging over a long period ($\D\log(B/G)=0.5$).} \label{rBaver}
\end{center}
\end{figure}

%%%%%%%%%%%%%%%%%%%%%%%%%%%%%%%%%%%%%%%%%%%%%%%%%%%%%%%%%%%%%%%%%%%%%%%
Figure.\ref{rB} shows the critical angular velocity $\O_c$ of a
strange quark star with mass $M=1.4 M_{\odot}$ and radius $R=10$ km
as a function of the magnetic field $B$ near $10^{18}$ G. The
temperature is fixed as $T=0.001$ MeV and other parameters are taken
according to \eq{paras}. In obtaining \fig{rB}, we have taken into
account the thermodynamical and hydrodynamical stability conditions.
The blue-solid curves correspond to the thermodynamically and
hydrodynamically stable region, while the red-dashed curves
correspond to unstable regions. The critical angular velocity is
strongly oscillating with increasing $B$. This behavior is due to
the oscillating nature of the bulk viscosities $\zpara$ and $\zperp$
as shown in \fig{zeta_u}. Thus, this macroscopic behavior originates
from a purely quantum mechanical effect, namely the Landau
quantization of the energy levels of quarks. As discussed for the
bulk viscosities $\zpara$ and $\zperp$ shown in \fig{zetaaver},
averaging is needed to obtain physically relevant quantities. In
\fig{rBaver} we show the averaged critical angular velocity  at
various temperatures.The solid black curves are obtained by
averaging over a short period $\D\log(B/G)=0.05$, whereas the
short-dashed red curves correspond to averaging over a long period
$\D\log(B/G)=0.5$. It is seen that after short-period averaging, the
critical angular velocity (solid black curves) shows regular
oscillation, the amplitude of which is growing as the $B$-field
increases. The critical angular velocity $\O_c$ displays a sharp
drop for fields $B\le 10^{18.5}$ G (short-dashed red curves), which
is the consequence of the sharp drop of $\zpara$ and $\zperp$ shown
in \fig{zetaaver}. Thus, we conclude that for extremely large
magnetic fields, the critical angular velocity at which the r-mode
instability sets in could be significantly lower than in the absence
of magnetic field.
%%%%%%%%%%%%%%%%%%%%%%%%%%%%%%%%%%%%%%%%%%%%%%%%%%%%%%%%%%%%%%%%%%%%%%%
%%%%%%%%%%%%%%%%%%%%%%%%%%%%%%%%%%%%%%%%%%%%%%%%%%%%%%%%%%%%%%%%%%%%%%%

\begin{figure}[!htb]
\begin{center}
\includegraphics[width=8.9cm]{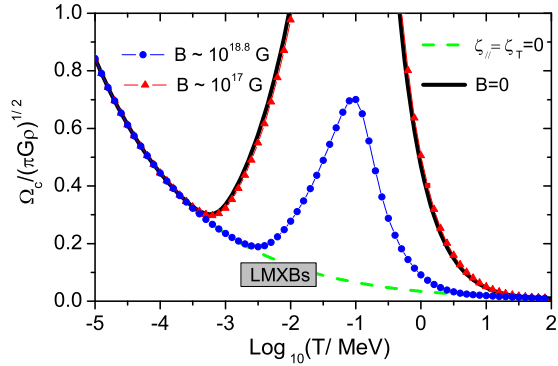}
\caption{(Color online) The r-mode instability window for a strange
quark star. The star is stable below the respective curves. The
green dashed curve corresponds to vanishing bulk viscosities
$\zpara=\zperp=0$. The black solid curve represents the window of
unmagnetized strange quark matter. The curves with symbols show the
typical behavior of the instability window when the magnetic fields
are around $10^{17}$ G (red curve) and $10^{18.8}$ G (blue curve).
The shadowed box represents typical temperatures ($2\times
10^7-3\times10^8$ K) and rotation frequencies (300-700 Hz) of the
majority of observed LMXBs \cite{Brown:2002}.} \label{rmode}
\end{center}
\end{figure}

Figure.\ref{rmode} shows the window of the r-mode instability in the
$\O-\log_{10}T$ plane for a strange quark star of mass $M=1.4
M_{\odot}$ and radius $R=10$ km. The regions above the respective
curves correspond to the parameter space where the r-mode
oscillations are unstable, \ie, a star in this region will rapidly
spin down by emission of gravitational waves. The dashed green curve
corresponds to vanishing bulk viscosities $\zpara=\zperp=0$. The
solid black curve represents the (in)stability window of an
un-magnetized strange quark star. The curves with symbols show the
typical instability window for magnetic fields around $10^{17}$ G
(the red curve marked by triangles) and $10^{18.8}$ G (the blue
curve, marked by circles). The symbolled curves are obtained by
using the bulk viscosities averaged over the period $\D
\log_{10}(B/G)=0.5$. For low temperatures, $T<0.3$ keV, the r-mode
instability is suppressed mainly by the shear viscosity; at these
low temperatures the bulk viscosities are an insignificant source of
damping, independent of how large the magnetic field is. However,
for larger temperatures the bulk viscosities dominate the damping of
r-mode oscillations. For magnetic fields below $B\sim10^{17}$ G, the
critical rotation frequency is almost independent of the $B$-field.
The r-mode instability window increases as the magnetic field grows.
For fields $B>10^{18}$ G it is a very sensitive function of the
field, as a consequence of the rapid variation of the bulk
viscosities with the field. Asymptotically, the instability window
can become significantly larger than the window at zero magnetic
field (see also \fig{rBaver}). For completeness, \fig{rmode} also
shows the observed distribution of Low Mass X-ray Binaries (LMXBs)
by the shadowed box, which corresponds to the typical temperatures
($2\times 10^7-3\times10^8$ K) and rotation frequencies (300-700 Hz)
of the majority of observed LMXBs \cite{Brown:2002}. It is seen that
even in the case of extremely large magnetic fields, our instability
window is consistent with the current LMXB data.

%%%%%%%%%%%%%%%%%%%%%%%%%%%%%%%%%%%%%%%%%%%%%%%%%%%%%%%%%%%%%%%%%%%%%%%
%%%%%%%%%%%%%%%%%%%%%%%%%%%%%%%%%%%%%%%%%%%%%%%%%%%%%%%%%%%%%%%%%%%%%%%

%%%%%%%%%%%%%%%%%%%%%%%%%%%%%%%%%%%%%%%%%%%%%%%%%%%%%%%%%%%%%%%%%%%%%%%
\section {Summary}\label{7}
%%%%%%%%%%%%%%%%%%%%%%%%%%%%%%%%%%%%%%%%%%%%%%%%%%%%%%%%%%%%%%%%%%%%%%%
In this paper we have studied anisotropic hydrodynamics of strongly
magnetized matter in compact stars. We find that there are in
general eight viscosity coefficients, six of them are identified as
shear viscosities, the other two, $\zperp$ and $\zpara$, are bulk
viscosities [see \eq{pmn}]. We applied our formalism to magnetized
strange quark matter and gave explicit expressions for the bulk
viscosities [\eq{vis1} and \eq{vis2}] due to the non-leptonic weak
reactions (\ref{weak1}) and (\ref{weak2}). Due to the Landau
quantization of the energy levels of charged particles in a strong
magnetic field, the magnetic field dependence of $\zpara$ and
$\zperp$ is very complicated and exhibits ``quasi-periodic" de Haas
- van Alphen type oscillations (see \fig{zeta_u}). For a magnetic
field $B\le 10^{17}$~G the effect of the magnetic field on the
transport coefficients is small and the bulk viscosities can be well
approximated by their zero-field values. For large fields $
10^{17}\le B \le 10^{20}$~G the viscosities are substantially
modified, $\zperp$ may even become negative for some values of the
$B$-field. We showed that  negative $\zperp$ render the fluid
hydrodynamically unstable.

For a number of reasons (density dependence of parameters along the
star profile, formation of domains, intrinsic multicomponent nature
of the magnetic field) the dependence of the transport coefficients
on the magnetic fields are needed at different resolutions, \ie,
they require some suitable averaging over a range of magnetic field.
We have provided such averages over an increasingly larger scale. We
find that if the averaging period is small the bulk viscosities show
regular oscillations, the amplitudes of which increase with magnetic
field (see the black solid curves in \fig{zetaaver}). These
oscillations are smoothed out if we further increase the averaging
scale. At this larger scale the most interesting feature is the
rapid drop in the bulk viscosity of the matter due to the
confinement of quarks to the lowest Landau level; this occurs for
magnetic fields in excess of $B>10^{18.5}$ G (see the short-dashed
red curves in \fig{zetaaver}).

As an application, we utilized our computed anisotropic bulk
viscosities to study the problem of damping of r-mode oscillations
in rotating Newtonian stars. We find that the instability window
increases as the magnetic field is increased above the value $B>
10^{17}$ G. By increasing the field one covers the entire range of
parameter space which lies between the two extremes: the case when
bulk viscosity vanishes (dashed green curve in \fig{rmode}, which
corresponds to extremely large magnetic fields $B\gtrsim10^{19}$ G
for which the bulk viscosity drops to zero), and the case when the
magnetic field is absent (solid black curve in \fig{rmode}). The
found novel dependence of the r-mode instability window on the
magnetic field may help to distinguish quark stars from ordinary
neutron stars with strong magnetic fields, since the latter are much
more difficult to magnetize. It would be interesting to see whether
the objects that lie in between these extremes, \eg, hybrid
configurations featuring quark cores and hadronic envelopes ~(see
ref.~\cite{Ippolito:2007hn} and references therein), may interpolate
smoothly between the physics of ordinary and strange compact
objects.

\section*{Acknowledgments} We thank T.~Brauner, T.~Koide, B.~Sa'd,
A.~Schmitt, and I.~Shovkovy  for helpful discussions. This work is
supported, in part, by the Helmholtz Alliance Program of the
Helmholtz Association, contract HA216/EMMI ``Extremes of Density and
Temperature: Cosmic Matter in the Laboratory'' and the Helmholtz
International Center for FAIR within the framework of the LOEWE
(Landesoffensive zur Entwicklung Wissenschaftlich-\"Okonomischer
Exzellenz) program launched by the State of Hesse. M.~H. is
supported by CAS program ``Outstanding young scientists abroad
brought-in", CAS key project KJCX3-SYW-N2, NSFC10735040,
NSFC10875134, and by K.~C.~Wong Education Foundation, Hong Kong.

%%%%%%%%%%
\appendix
%%%%%%%%%%

%%%%%%%%%%%%%%%%%%%%%%%%%%%%%%%%%%%%%%%%%%%%%%%%%%%%%%%%%%%%%%%%%%%%%%%
\section{Linear independence of components in \eq{combination}} \label{proof}
%%%%%%%%%%%%%%%%%%%%%%%%%%%%%%%%%%%%%%%%%%%%%%%%%%%%%%%%%%%%%%%%%%%%%%%
The purpose of this appendix is to show explicitly  that the
eight different decompositions in \eq{combination}
are linearly independent. To this end, let us write
down a general linear combination of the eight different
decompositions,
\begin{eqnarray}
\label{linear0} a_1({\rm i})+a_2 ({\rm ii})+\cdots+a_8 ({\rm
viii})=0.
\end{eqnarray}
If $({\rm i})-({\rm viii})$ are linearly independent, the
coefficients $a_1-a_8$ should all vanish for any values of the
vectors $u^\m$ and $b^\m$. Firstly, it is easy to see that the
components $({\rm vi})$ and $({\rm vii})$ are independent of the
other components, because they have odd parity under reflection
$b^\m\ra-b^\m$, whereas the other six components have even parity
under this transformation. Besides that it is obvious that $({\rm
vi})$ and $({\rm vii})$ are independent of each other. Therefore, we
only need to treat the linear equation
\begin{eqnarray}
\label{linear2} a_1({\rm i})+\cdots+a_5({\rm v})+a_8 ({\rm viii})=0.
\end{eqnarray}
By contracting the indices
$\m$ and $\n$, we obtain the following three conditions
\begin{eqnarray}
\label{linear3}
&3a_1+2a_2-a_3+2a_8 = 0,&\non
&3a_3-a_4+4a_5+2a_8 = 0,&\non
&3a_1+2a_2-12a_3+a_4-4a_5=0.&
\end{eqnarray}
Contracting \eq{linear0} with $b^\m$ and $b^\n$ we find the
following two additional conditions
\begin{eqnarray}
\label{linear4}
&a_1+a_3 = 0,&\non
&2a_2-a_3+a_4-4a_5 = 0.&
\end{eqnarray}
Contracting the indices $\n$ and $\a$ in \eq{linear}, we obtain one
further condition
\begin{eqnarray}
\label{linear5}
&a_1+4a_2-2a_3+a_4-6a_5 = 0.&
\end{eqnarray}
The non-trivial solution of the set of \eqs{linear2}{linear5} is
\begin{eqnarray}
\label{linear6}
&a_1=a_3=0,&\non
&a_2=a_4/2=a_5=-a_8.&
\end{eqnarray}
Then, we have the following condition,
\begin{eqnarray}
\label{linear7}
a_2(b^{\m\a}b^{\n\b}+b^{\n\a}b^{\m\b}-\Xi^{\m\a}\Xi^{\n\b}-\Xi^{\n\a}\Xi^{\m\b})=0.
\end{eqnarray}
The only possible solution is $a_2=0$, which thus proves the
independence of $({\rm i})-({\rm viii})$.

\end{document}